\documentclass[11pt]{article}
\usepackage{fullpage,amsthm}
\usepackage{amsmath,amssymb,amsfonts}
\usepackage{algorithmic}
\usepackage{graphicx}
\usepackage{textcomp}
\usepackage{xcolor}
\usepackage{longtable}
\usepackage{indentfirst}
\usepackage{subcaption}
\usepackage{hyperref}
\usepackage[style=ieee]{biblatex}
\DeclareMathOperator{\EX}{\mathbb{E}}

\begin{document}
\baselineskip 12pt

\begin{center}
\textbf{\Large Numerical Model Simulation of the Carruthers GCI Images} \\

\vspace{1.5cc}
{ \sc Heather Filippini$^{*1}$, Alex Zhang$^{2}$, Evan Widloski$^{2}$, Jason McPhate$^{3}$, Lara Waldrop$^{2}$, Thomas Immel$^{3}$, John Clarke$^{4}$, Pratik Joshi$^{2}$, Michal Ondrejcek$^{1}$, Martin M. Sirk$^{3}$}\\

\vspace{0.3 cm}

{\small $^{1}$Illinois Coordinated Sciences Laboratory, University of Illinois Urbana-Champaign \\
$^{2}$Department of Electrical and Computer Engineering, University of Illinois Urbana-Champaign \\
$^{3}$ Space Sciences Laboratory, University of California, Berkeley \\
$^{4}$ Department of Astronomy and Center for Space Physics, Boston University \\
}
 \vspace{0.3 cm}
{\small $^{*}$Corresponding Author: hrf@illinois.edu}
 \end{center}

 \vspace{1.5cc}

\begin{abstract}
  \noindent 
  The Carruthers Geocorona Observatory, launched in September 2025, is NASA’s first mission devoted to investigating the fundamental nature of Earth’s exosphere from its distant vantage in halo orbit around the Earth-Sun Lagrange 1 (L1) point. Its primary payload, the GeoCoronal Imager, consists of two coaligned photometric imagers that measure the radiance of ultraviolet emission at 121.6 nm (Lyman-$\alpha$, or Ly-$\alpha$) from exospheric hydrogen atoms simultaneously at wide and narrow fields of view. In order to validate the calibration and hydrogen density retrieval algorithms used in the Carruthers data processing pipeline, we developed a comprehensive numerical simulator to produce realistic images similar to those collected by the actual imagers on orbit. This paper details the algorithms used to simulate the exospheric emissions, background scene components, and instrument measurement model necessary to produce synthetic raw images. 
\vspace{0.95cc}

\parbox{24cc}{{\it Key words and Phrases}: Exosphere, Carruthers Geocorona Observatory, GCI instrument, Simulation
}
\end{abstract}

\section{Introduction}

The Carruthers Geocorona Observatory is a NASA Heliophysics Mission of Opportunity to investigate the structure and dynamics of Earth's exosphere by imaging the  ultraviolet emission by its constituent hydrogen (H) atoms at 121.6 nm (Lyman-$\alpha$, or Ly-$\alpha$).  The Carruthers mission launched in September, 2025, into halo orbit around the Earth-Sun L1 Lagrange equilibrium point, where it has an unobstructed, wide-field view of exospheric Ly-$\alpha$ emission and full spacecraft attitude control to support routine slewing off of Earth nadir for calibration data acquisition.  From this ideal vantage for exospheric remote sensing, Carruthers will measure the exospheric Ly-$\alpha$ radiance distribution with unprecedented spatial resolution and temporal cadence in support of the mission's science goal to determine the nature and origin of observed spatial asymmetries and transient variability in the exospheric H density distribution.    

The Carruthers Geocorona Observatory mission's primary scientific payload is the GeoCoronal Imager (GCI), which consists of two co-aligned imagers (channels) for simultaneous sensing of the exospheric Ly-$\alpha$ radiance distribution. The Narrow Field Imager (NFI) provides high spatial resolution within its 3.6$^{\circ}$ field-of-view (FOV), where gradients in exospheric radiance are expected to be largest. The Wide Field Imager (WFI) captures the nadir scene at lower spatial resolution within its larger, 18$^{\circ}$ FOV but at a sensitivity sufficient to detect dim Ly-$\alpha$ signals from the distant outer region of the exosphere. Both channels employ identical UV-intensified, Active Pixel Sensor (APS)-based cameras with heritage from the ICON mission's FUV imager \cite{mende2017iconfuvinstrument} along with identical 6-position filter wheels.  One filter position is unoccupied for open system transmission, one contains a solid aluminum block for dark calibration, and the other four contain optical filters for transmission or suppression of the target Ly-$\alpha$ emission. A detailed description of the GCI payload is given by \cite[]{sirk2026}.

Figure 1 illustrates the path of photons through one of the imaging channels, from their entry through the aperture to their measurement as integrated signal in digital numbers (DN).  The incident photons, including those comprising the in-band and out-of-band scene background, are collimated by curved mirrors coated with MgF2, transit one of the filter wheel openings and a 3-mm thick MgF2 tube window with 95\% Ni transmissive backside coating, and impinge on a Potassium Bromide (KBr) photocathode. Photon detection by the cathode liberates an energetic photoelectron, which is referred to as a photoelectron ``event''. The product of the window transmissivity, mirror reflectivities, optical filter transmissivity (if applicable), and KBr quantum efficiency constitutes the ``optical efficiency'', which has units of [photoelectron events/incident photon], or [events/photon] for short, and represents the fraction of incident photons that are detected. Each photoelectron is accelerated through a high voltage drop onto a microchannel plate (MCP), which produces a cloud of secondary electrons that impinge on a phosphor screen.  Phosphor detection of MCP electrons releases visible wavelength photons, which are fed through a fiberoptic taper bonded onto the APS detector, where they are detected as electron ``counts''.  Ambient cosmic rays or solar energetic particles in the spacecraft environment, termed ``particle radiation'' in Figure 1, interact with both the MCP and the APS detector, releasing secondary electrons that are detected as a transient background signal along with the more temporally stable but temperature dependent dark current inherent to the detector. Detector electron counts from both photon and non-photon sources accumulate in the native, non-dark, 2048x2048 detector pixel wells and are binned and read out by the camera Field Programmable Gate Array (FPGA) using a gain-amplified Analog-to-Digital Converter (ADC) in order to obtain the final instrument units of Digital Numbers [DN].  At camera read-out, a voltage bias is applied to each ADC to ensure that signals exceed the read noise level.  Individual frames of signals in DN are then co-added onboard before telemetry. The two imaging channels are nearly identical yet completely independent optical systems which differ by design only in terms of their fields-of-view, number of collimating mirrors along the optical path, and image binning and frame stacking settings.

\begin{figure}[ht]
\begin{center}
    \includegraphics[width=\textwidth]{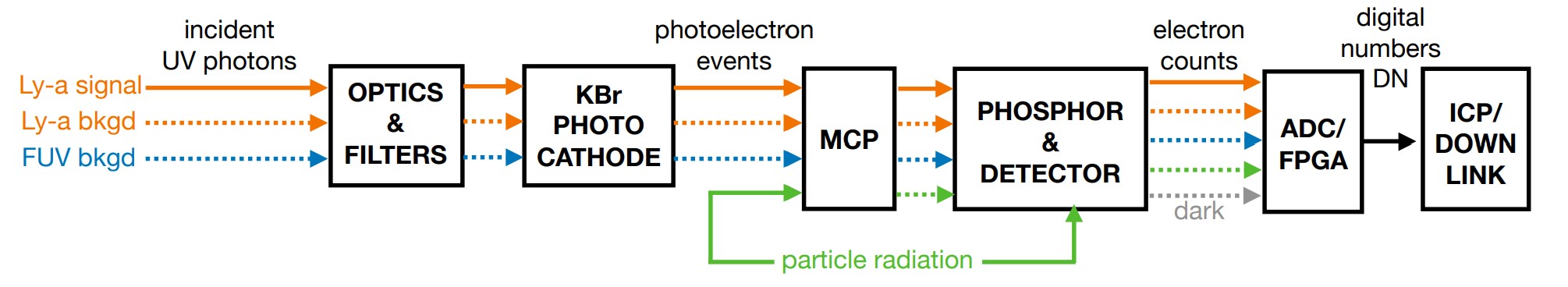}
    \caption{An overview of the GeoCoronal Imager measurement concept, illustrating the detection of incident target exospheric Lyman-alpha photons and their subsequent telemetry as digital numbers (DN).  Sources of background signals introduced by the scene (in-band and out-of-band photons), the spacecraft environment (energetic particle radiation), and the detector (dark current) are also shown.}
    \label{fig:gci_instrument_overview}
\end{center}
\end{figure}

Calibration of GCI images involves removing instrument effects such as dark current and optical distortion \cite[]{Zhang26c}, isolating the target exospheric signal from the photon background scene \cite[]{Zhang26a}, and converting the measured signals into physical units of exospheric Lyman-$\alpha$ emission radiance \cite[]{Zhang26b}.  Retrieval of the underlying exospheric H density involves reconciling the measured radiance distribution with photon scattering models of the target scene in terms of a parameterized H density distribution \cite[]{joshi26, widloski26}.  
Assessing the performance (accuracy and precision) of the image calibration and parameter retrieval algorithms against mission requirements requires realistic model simulation of both the expected scene and its observation by the GCI, including background specification and measurement noise.  

This paper defines the forward GCI image model, which is split into two independent model components: a scene model, which defines the physics-based transformation that maps the input physical parameters (e.g., assumed ground-truth exospheric H density) to a spectral radiance field, and an instrument model, which maps the spectral radiance field to the GCI’s raw output signals. Next, this paper develops a numerical image simulator that implements both the scene model and the instrument model in a computationally efficient manner. Other papers in this special issue heavily utilize this numerical image simulator to support Monte Carlo validations of image calibration and H density retrieval algorithms used in the Carruthers mission data processing pipeline.

\subsection{Notation and Units}

This paper uses the following conventions for notation:
\begin{itemize}
    \item Matrices, sets, and lists will be denoted as upper-case and bold-font English letters, such as $\boldsymbol{X}$ or $\boldsymbol{Y}$.
    \item Random variables will be denoted as upper-case English letters, such as $X$ or $Y$.
    \item Vectors will be denoted as lower-case English letters or Greek letters with an arrow above the symbol, such as $\vec{x}$ or $\vec{y}$.
    \item Estimated quantities will be designated with a hat on top of the symbol for the corresponding quantity. For example, $\hat{x}$ is the estimated value for $x$.
    \item All other functions and constants will be denoted as lower-case English or Greek letters.
\end{itemize}

Variables $x$ that are a function of another variable $y$ are written explicitly as $x(y)$ to avoid confusion; due to the number of symbols required, some symbols are overloaded. Thus, a constant $x$ can have a completely different meaning compared to the function $x(\lambda)$, even though both are labeled using the same letter. In some cases, names are placed as a subscript for clarity, such as $x_{\text{example}}$ or $y_{\text{Ly-}\alpha}$.

Units are explicitly enclosed in brackets (e.g., $[\text{cm}]$). Spectral radiances are calculated in the conventional physical unit of $[10^6 \; \text{phot}/\text{s}/\text{cm}^2/\AA]$, which is formally defined as the apparent column emission rate at a given wavelength assuming isotropic radiation over a full sphere. Wavelength-integrated emission radiances, used to quantify line emissions associated with discrete electronic transitions as opposed to continuum excitation processes, are calculated in the conventional physical unit of $[10^6 \; \text{phot}/\text{s}/\text{cm}^2]$.  We avoid the historically ambiguous term of ``Rayleigh" to describe this radiance unit (see Baker and Romick (1976) \cite{baker1976rayleigh}), since we explicitly do not include the optional normalization of the assumed isotropic emission over $4\pi$ [sr]   \cite{hunten1956photometric}.

\section{Viewing Geometry}
This section describes the viewing geometry of the Carruthers observatory and the projection geometry of the NFI and WFI cameras. The nominal Carruthers ephemeris is a halo orbit with a six-month period around the Earth-Sun L1 Lagrange point.
The ephemeris is defined in the EME J2000 world coordinate frame, also known as the Geocentric Celestial Reference System (GCRS). This coordinate frame is an inertial Earth-centric frame and uses the J2000 epoch.

\begin{figure}
    \begin{subfigure}{0.3\textwidth}
      \includegraphics[height=2.0in]{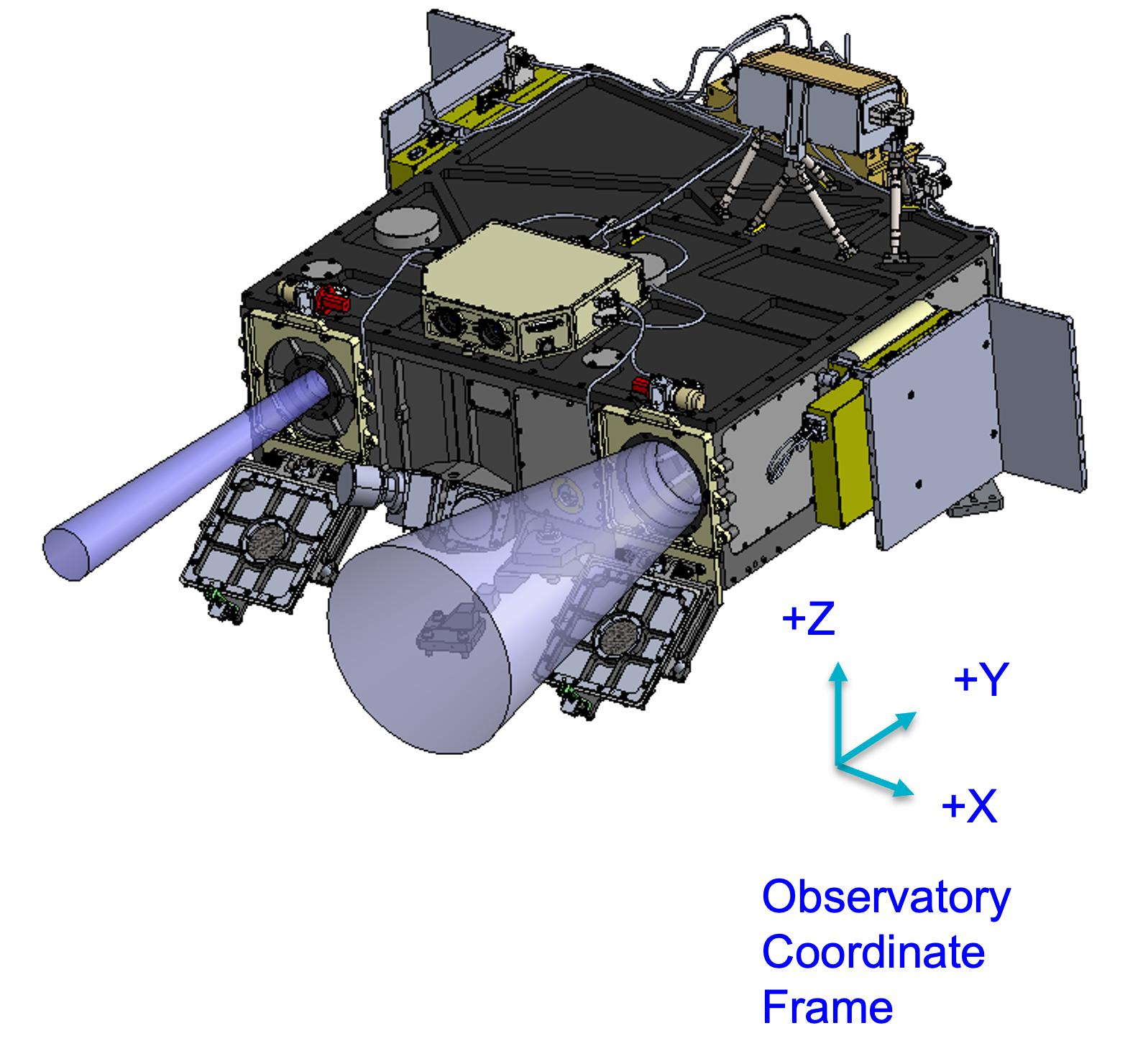}
      \caption{Observatory coordinate frame definition.}
      \label{fig:Observatory_Coordinates}
    \end{subfigure}
    \hfill
    \begin{subfigure}{0.3\textwidth}
      \includegraphics[height=2.2in]{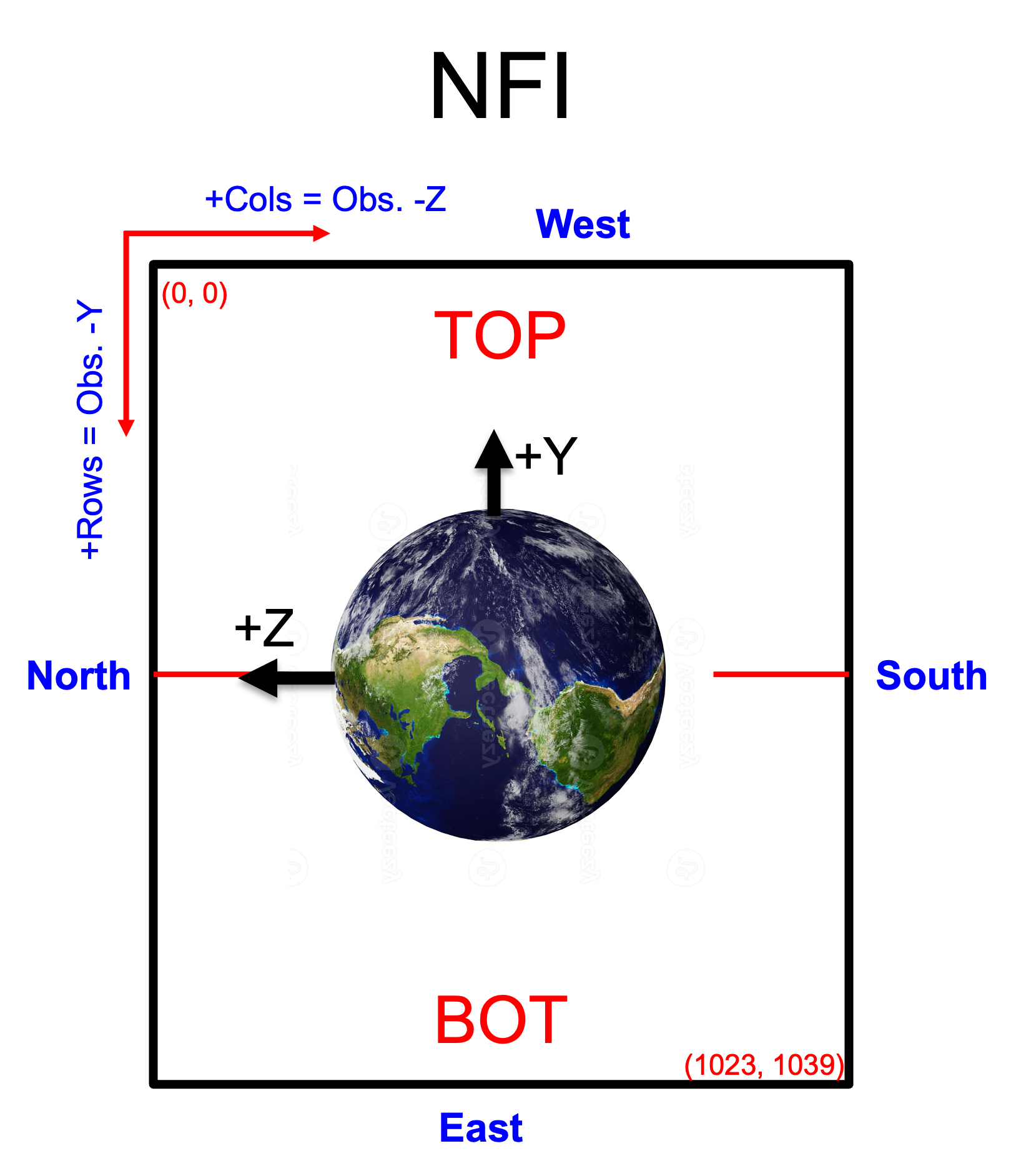}
      \caption{Projection of Earth and GSE y and z axes in NFI coordinates for the spacecraft positioned along the Earth-Sun line.}
      \label{fig:NFI_L0_coordinates}
    \end{subfigure}
    \hfill
    \begin{subfigure}{0.3\textwidth}
      \includegraphics[height=2.2in]{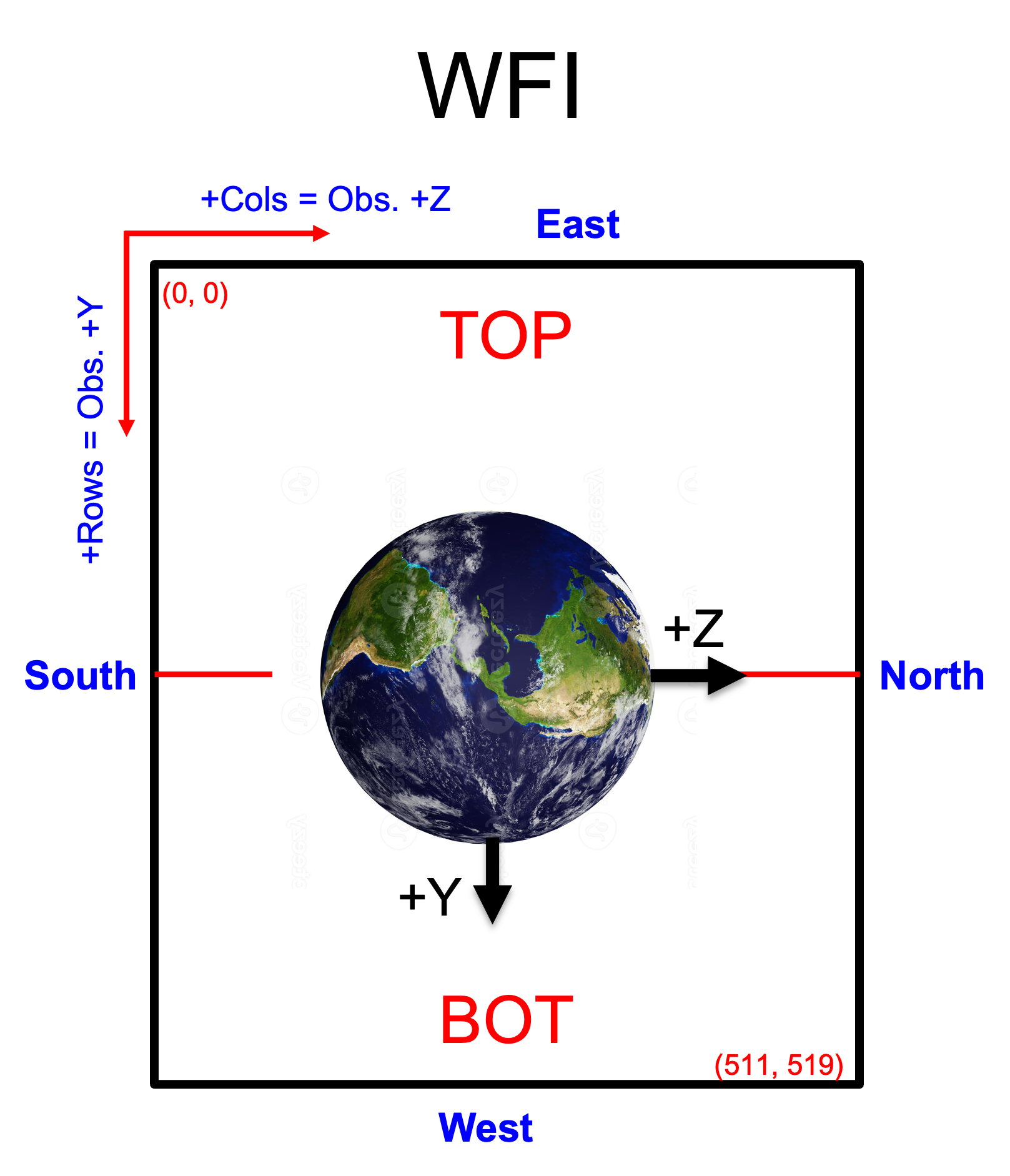}
      \caption{Projection of Earth and GSE y and z axes in WFI coordinates for the spacecraft positioned along the Earth-Sun line.}
      \label{fig:WFI_L0_coordinates}
    \end{subfigure}
    \hfill
    \caption{}
\end{figure}

The observatory is constructed with the fixed solar panels oriented opposite to the GCI imagers' co-aligned boresights, which enables continuous, power positive, nadir-staring. The two imagers also have fixed orientations with respect to the observatory, with their boresights co-aligned with the observatory -$\vec{y}$ axis as seen in Figure \ref{fig:Observatory_Coordinates}. However, the two imagers' optical paths are independent and, due to their respective read-out orientations, the NFI and WFI images are rotated 180 degrees with respect to each other. Figure \ref{fig:NFI_L0_coordinates} - \ref{fig:WFI_L0_coordinates} depicts the projected view of Earth in the NFI and WFI images planes, respectively, if the spacecraft is positioned along the Earth-Sun line with its +$\vec{z}$ axis aligned with Earth North. To simulate the viewing geometry for both on- and off-nadir images, the observatory position is set using either the predictive ephemeris (pre-launch) or the definitive on-orbit ephemeris. For on-nadir images, the observatory attitude is then adjusted such that the imagers' boresights point at Earth. For off-nadir images, the observatory attitude is adjusted to point the imagers' boresights at the desired celestial right-ascension and declination. 

The total coordinate transform from world coordinates to imager coordinates is defined as shown in Equation \ref{eq:world_transform}, where $\boldsymbol{T}_{obs}$ represents the rotation and translation between the GCRS world coordinate frame and the observatory coordinate frame and $\boldsymbol{R}_{cam}$ represents the rotation between the observatory coordinate frame and the imager coordinate frame. The translational offset between the imagers and the observatory is negligible compared to the distance of the objects being observed and can therefore be safely omitted. 

\begin{equation}
    \label{eq:world_transform}
    \vec{x}_{cam}=\boldsymbol{R}_{cam}\boldsymbol{T}_{obs}\vec{x}_{world}
\end{equation}

Each imager is represented by a pinhole camera model. Equation \ref{eq:proj_eq} maps the homogeneous imager coordinates to the imager focal plane using the intrinsic calibration matrix $\boldsymbol{K}$. Using these equations, every element of the scene model can be projected from world coordinates to the imagers' focal planes. To simplify modeling of the scene, optical distortion is applied to the composite photon scene as part of the instrument model as described in Section \ref{sec:instrument_model}.

\begin{equation}
    \label{eq:proj_eq}
    \vec{u}=\boldsymbol{K}\frac{\vec{x}_{cam}}{z_{cam}}
\end{equation}

\section{Scene Model}
\label{sec:scene_model}

This section details the scene model for the incident omnidirectional spectral radiance $\ell(\lambda, i, j)$ as a function of wavelength $\lambda$, pixel row index $i$, and pixel column index $j$, in units of [photons/s/cm$^2$/{\AA}]. The scene is synthesized by independently modeling and aggregating four distinct photon sources: the science target terrestrial exospheric Ly-$\alpha$, non-terrestrial Ly-$\alpha$, the terrestrial Out-Of-Band (OOB) background spectrum, and celestial UltraViolet (UV) photon sources (such as the stars, the Moon, and the outer planets).

\subsection{Exospheric \texorpdfstring{Ly-$\alpha$}{Ly-alpha} Scene Model}
\label{sec:exospheric_scene_model}

Exospheric Ly-$\alpha$ emission, the target observable of the Carruthers mission, is generated by the resonant scattering of solar Ly-$\alpha$ photons by exospheric H atoms \cite[]{meier_1991}.  In the inner exosphere, nominally within 3 Re from Earth's surface, the H density is sufficiently large that Ly-$\alpha$ photons experience multiple resonant scattering interactions among ambient H atoms before their eventual detection.  In this optically thick emission regime, modeling the observed Ly-$\alpha$ emission radiance, $i_{1216}(i, j)$ [phot/s/cm$^2$], in terms of the density distribution of the scattering H population requires the use of a radiative transfer (RT) model, which must account for the spherical geometry of the extended scattering region, thermal broadening of the scattered emission line, and non-isothermality throughout the thermosphere.  The RT model used for the simulation of the GCI scene is based on the LYAO-RT code developed by \cite{bishop1999lyao_rt_code}, modified as described by \cite[]{qin2016}.  
Implementation of the LYAO-RT model in the GCI image simulator requires \emph{a priori} specification of a parameterized H density distribution throughout the exosphere, and two options based on analytic theory are supported: (1) the classical one-dimensional (1D) Chamberlain model of thermal evaporation \cite{chamberlain1963}, and (2) a modification of the Chamberlain model that accounts for a non-negligible satellite atom population \cite[]{bishop1991}.  Both of these exospheric H density models require the extension of the H density distribution into the thermosphere, and the simulator adopts the approach described in \cite{bishop2001} using background atmosphere specification by the MSIS 2.0 model \cite[]{emmert2021}.   

In the outer exosphere beyond $\sim$3 Re, where exospheric Ly-$\alpha$ emission is optically thin, the emission radiance is linearly proportional to the integrated density of H atom scatterers along the viewing column.  As a result, the scene model can be simulated straightforwardly by ray-tracing the GCI lines-of-sight through an exospheric H density distribution specified on a spherical grid of arbitrary radial extent.
The H density model in this case also can be arbitrary, and several options for parameterized specification based on the analytical Chamberlain model or data-driven spherical harmonic models (e.g., \cite[]{zoennchen2024}) are supported by the simulator.  

No unified Ly-$\alpha$ emission model is currently available that produces a realistic two-dimensional (2D) scene across both the inner and outer exospheric regions.  The optically thin outer exospheric emission model is capable of supporting spherically asymmetric density distributions, and thus can produce azimuthally asymmetric radiance distributions in the simulated image, but its implicit neglect of multiple scattering renders it inappropriate for simulating radiances along lines-of-sight that transit the inner exosphere and especially the Earth's limb, where optical depth is largest.  Meanwhile, the computational complexity of the LYAO-RT model of multiple photon scattering in the optically thick inner exosphere requires formulating the H density distribution in terms of a simplified, 1D profile that varies only with radial distance.  Moreover, confidence in the validity of the analytical formulations of the 1D profile supported by the LYAO-RT model is highest in the inner exosphere only, where their assumptions regarding H population thermalization are most likely to be valid.       

\begin{figure}[htbp]
    \centering
    \includegraphics[width=1.0\linewidth]{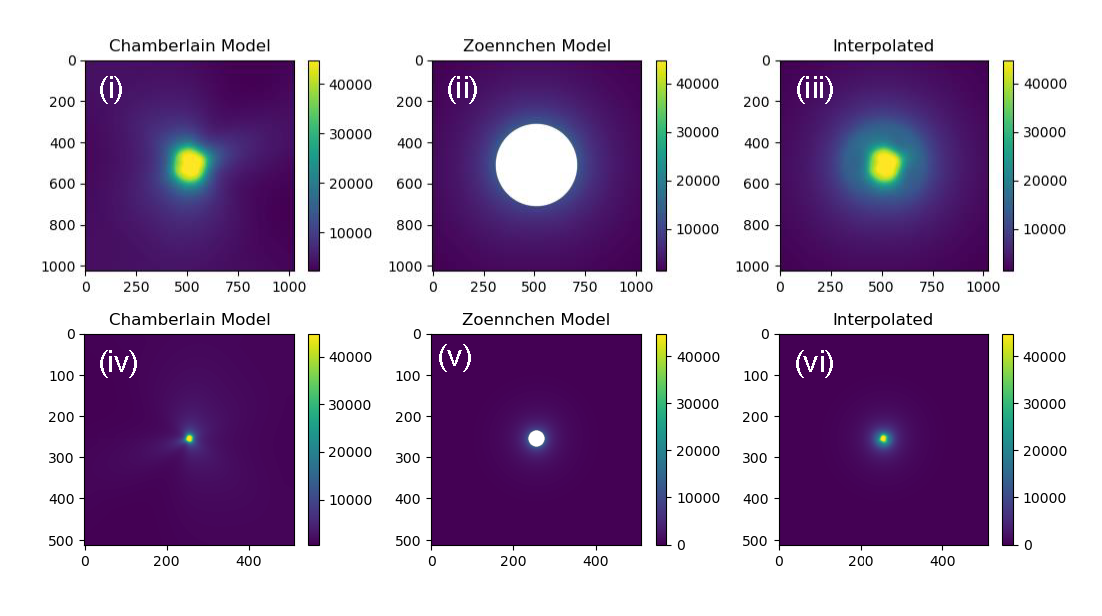}
    \caption{Exospheric scene models used as input to the GCI image simulator. Panels (i)-(iii) depict the 2D scene radiance in $[10^6 \text{phot}/\text{s}/\text{cm}^2]$ as viewed by the high resolution (1024x1024 pixel) Narrow Field Imager (NFI), while panels (iv)-(vi) depict those as viewed by the lower resolution (512x512) Wide Field Imager (WFI).  The azimuthally asymmetric inner exosphere radiance distribution (panels (i) and (iv)) is constructed as the azimuthal interpolation over independent LYAO-RT emission model runs using a Chamberlain formulation for the exospheric H density distribution.  The outer exospheric radiance distribution is calculated using ray-tracing through the spherically asymmetric H density distribution specified by \cite{zoennchen2015}. Panels (iii) and (vi) depict the final composite scene used for GCI image simulation and end-to-end algorithm validation.  This scene is calculated by interpolating over the independent inner and outer exospheric scenes.}
    \label{fig:exo_scene}
\end{figure}

These exospheric emission model limitations preclude the direct simulation of a realistic 2D radiance scene for input to the GCI image simulator.  To overcome this challenge, we construct a simulated Ly-$\alpha$ radiance scene as a composite of both emission models as shown in Figure ~\ref{fig:exo_scene}.  Specifically, the outer exosphere is simulated using the Zoencchen model \cite{zoennchen2015}.  For the inner exosphere out to 1.5 Re, we used the evaporative Chamberlain H density distribution to generate 1D, RT-modeled radial emission profiles independently along multiple line-of-sights arranged in a ring about Earth as viewed from the GCI vantage. We then interpolated between these profiles azimuthally in the radiance domain to produce a full, azimuthally-asymmetric, synthetic image of the inner exospheric radiance distribution.  We found that using 60 azimuthal profiles provided a good balance between computational cost and smoothly varying radiance. 
Lastly, we interpolated the independently modeled inner and outer exospheric radiance distributions across the image annulus spanning 1.5-3 Re.  While the resultant interpolated region is not necessarily physically realistic, none of the H density retrieval algorithms use image data in that region, so any errors are unlikely to significantly impact the validity of the retrieval performance assessments \cite[]{joshi26, widloski26}.

To convert the band-integrated intensity into exospheric omnidirectional spectral radiance $\ell_{\text{exo}}(\lambda, i, j)$ [phot/s/cm$^2$/{\AA}], the emission profile is approximated as a Dirac delta function centered at 1216{\AA}. The approximation is justified by the extremely narrow Doppler width of the geocoronal Ly-$\alpha$ line ($\sim$0.02{\AA} at typical exospheric temperatures of 1000K) \cite{chamberlain1990theory_thin_width}, which is negligible compared to the spectral scale of the system efficiency features \cite{sirk26a}. Consequently, the exospheric Ly-$\alpha$ emission can be treated as effectively monochromatic.

\begin{equation}
    \label{eq:l_exo_def}
    \ell_{\text{exo}}(\lambda, i, j) = i_{1216}(i, j)\delta(\lambda - 1216)
\end{equation}

\subsection{Non-terrestrial \texorpdfstring{Ly-$\alpha$}{Ly-alpha} Scene Model}

\begin{figure}[htbp]
    \centering
    \includegraphics[width=0.8\linewidth]{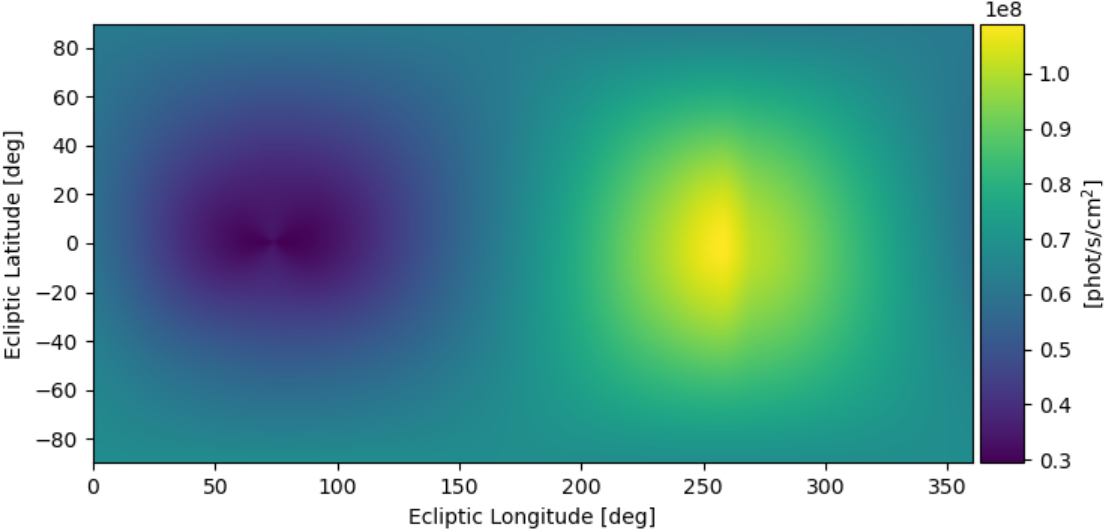}
    \caption{Example IPH Ly-$\alpha$ intensity derived using the Pryor model for solar maximum conditions. The IPH Ly-$\alpha$ intensity is an in-band background of the target exospheric Ly-$\alpha$ emission. The axes use the Heliocentric Mean Ecliptic coordinate system.}
    \label{fig:iph_map}
\end{figure}

The in-band Ly-$\alpha$ photon background is dominated by emission from InterPlanetary Hydrogen (IPH) atoms, where the H atoms are of interstellar origin. The in-band radiance resulting from IPH atoms is denoted $i_{\text{IPH}}(i, j)$ [phot/s/cm$^2$]. The IPH map, in ecliptic longitude and latitude, is produced by a physics-based model \cite{IPHModelPryor2013} and is shown in Figure \ref{fig:iph_map} in the Heliocentric Mean Ecliptic coordinate system.\footnote{The Heliocentric Mean Ecliptic coordinate system is a spherical coordinate system centered at the Sun with the mean ecliptic plane of J2000.0 as the reference plane. Longitude is defined as the angular coordinate in the ecliptic plane, measured from the vernal equinox, while latitude is defined as the angular distance of the object above or below the ecliptic plane. Radial distance can be optionally specified, but only the directional components are relevant for IPH.} Note that $i_{\text{IPH}}(i, j) = 0$ in areas of the image where the IPH is eclipsed by another object (such as the Earth or the Moon). To convert the IPH intensity into IPH omnidirectional spectral radiance $\ell_{\text{IPH}}(\lambda, i, j)$ [phot/s/cm$^2$/{\AA}], the emission profile is approximated as a Dirac delta function centered at 1216{\AA}. This approximation is justified because the IPH is also a narrowband line emission produced by solar resonant scattering \cite{bertaux1971iph_evidence}.
\begin{equation*}
    \ell_{\text{IPH}}(\lambda, i, j) = i_{\text{IPH}}(i, j)\delta(\lambda - 1216)
\end{equation*}

\subsection{Terrestrial Out-Of-Band Scene Model}

This section derives a model for $\ell_{\text{OOB}}(\lambda, i, j)$ [phot/s/cm$^2$/{\AA}], which denotes non-Ly-$\alpha$ (out-of-band, or OOB) omnidirectional spectral radiance. The exospheric Ly-$\alpha$ derived from radiative transfer models (as discussed in Joshi et al. 2026 \cite{joshi26}) can vary by an order of magnitude depending on solar/geomagnetic indices and observational geometry \cite{bishop1999lyao_rt_code}. In order to maintain one OOB spectral radiance model for any exospheric Ly-$\alpha$ input from the radiative transfer model, OOB spectral radiance is modeled as an analytic function of exospheric Ly-$\alpha$ radiance. In reality, no causal relationship between exospheric Ly-$\alpha$ radiance and OOB spectral radiances is known, but the OOB scene model maintains some realism since the total radiances share common physical drivers \cite{GUVI_info_and_initial_data}. Thus, the OOB scene model developed in this section is sufficient to create test cases for the calibration algorithms.

\begin{figure}[htbp]
    \centering
    \includegraphics[width=0.3\linewidth]{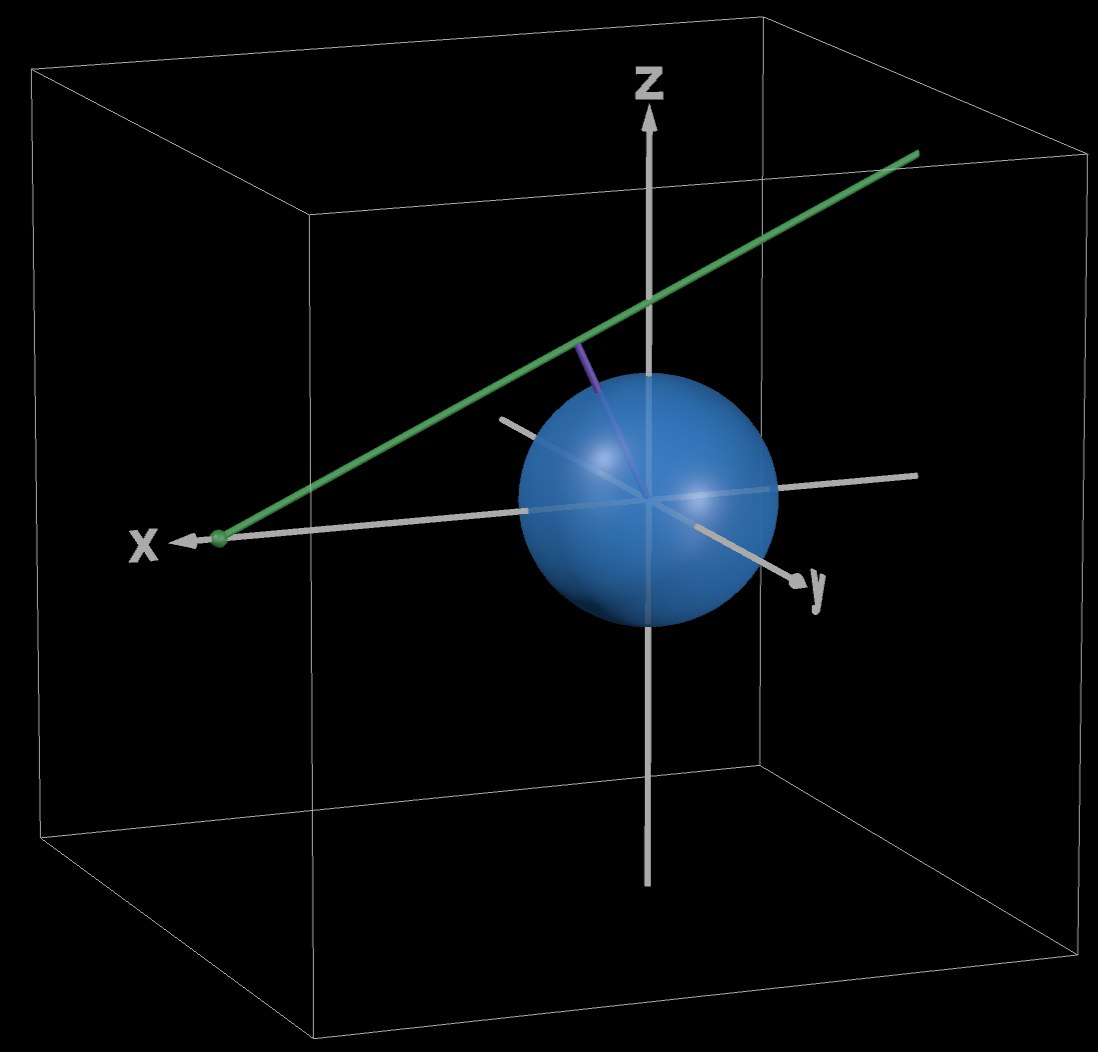}
    \caption{Example of $r_{ij}$, or altitude of tangent point of Line-of-Sight (LOS) to Earth. The blue ball is Earth, while the green dot is the spacecraft location. The green line is the LOS of some pixel $i,j$. The purple line is perpendicular to the green line; its length from the center of the Earth to the green line is the altitude $r_{ij}$. The size of Earth relative to the spacecraft location is not to scale.}
    \label{fig:los_tp_r_example}
\end{figure}

Throughout this section, let $r_{ij}$ denote the Euclidean distance between the center of the Earth and the tangent point of the line of sight (LOS) of each pixel $i,j$ to the surface of the Earth, in units of Earth-radii (Re). The tangent point is defined as the point of closest approach to the spherical body (see Figure \ref{fig:los_tp_r_example} for an example of $r_{ij}$). Let $\boldsymbol{D}$ be the set of pixels that satisfy $r_{ij} < 1$, which implies that the pixels in $\boldsymbol{D}$ span the Earth itself. The area of the image covered by the pixels in $\boldsymbol{D}$ is referred to as Earth's disk. Let $p_{ij}, q_{ij}$ be the pixel indices of the closest pixel in $\boldsymbol{D}$ to pixel $i, j$ (with respect to Euclidean distance on the image plane).

The three main OOB spectral radiance contributions that are relevant for the Carruthers mission are the two narrow Oxygen emission lines (at 1304{\AA} and 1356{\AA}) and the Nitrogen Lyman-Birge-Hopfield (LBH) emission band spanning 1400-1800{\AA} \cite{lbh_definition, far_ultraviolet_instrument_technology}. These emissions are modeled by first calculating the total radiance from each emission line $m$ and emission band $n$ as a function of image pixel $i,j$ based on data from the Global Ultraviolet Imager (GUVI) instrument aboard the Thermosphere, Ionosphere, Mesosphere Energetics and Dynamics (TIMED) spacecraft \cite{GUVI_info_and_initial_data}. GUVI measured the LBH band emission as separate LBHS (LBH-Short) and LBHL (LBH-Long) bands, so the OOB scene model also splits the LBH band emission into two spectral regions: one spanning 1400-1600{\AA} and one spanning 1600-1800{\AA}. Finally, the total radiance from each emission line $m$ and emission band $n$ is multiplied by the corresponding wavelength distribution function $g(\lambda)$ to obtain spectral radiance. In order to maintain the overall derived intensity, $g(\lambda)$ must satisfy $\int_0^{\infty} g(\lambda)d\lambda = 1$. For emission line $m$ centered at $\lambda_{m}$ (Oxygen lines), $g_{m}(\lambda) = \delta(\lambda - \lambda_{m})$. For emission band $n$ centered at $\lambda_n$ (Nitrogen bands), the OOB scene model assumes that the total radiance is equally spread over all wavelengths of the emission band for simplicity, resulting in $g_n(\lambda) = \frac{1}{200}\text{rect}\left(\frac{\lambda - \lambda_n}{200}\right)$.

\begin{figure}[htbp]
    \begin{subfigure}[t]{0.48\textwidth}
      \includegraphics[width=\textwidth]{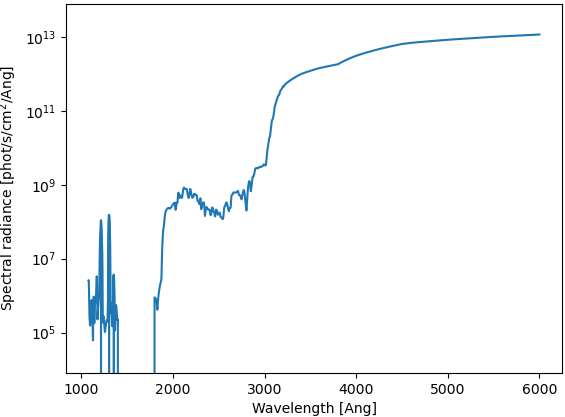}
      \caption{Earth disk spectrum with Hydrogen and Oxygen emission lines removed and Nitrogen emission bands removed. These emission lines and emission bands are modeled separately.}
      \label{fig:earth_disk_spectrum}
    \end{subfigure}
    \hfill
    \begin{subfigure}[t]{0.48\textwidth}
      \includegraphics[width=\textwidth]{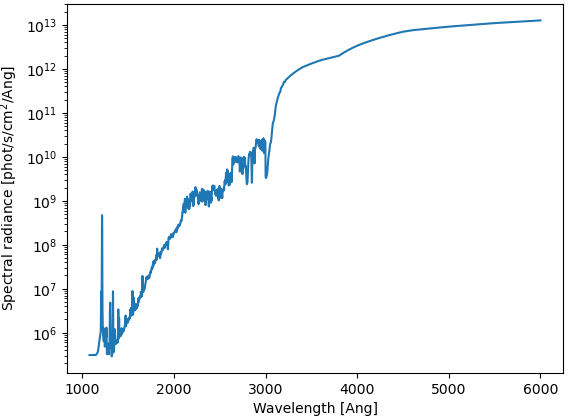}
      \caption{Lunar spectrum.}
      \label{fig:lunar_spectrum}
    \end{subfigure}
    \caption{Spectral radiance used for Earth disk and the Moon.}
\end{figure}

In the region $r_{ij} < 1$Re (Earth's disk), additional contributions from Earth's continuum emission are present in the measured photon signal. The spectral radiance of Earth's disk is obtained from Meier (1991) \cite{meier_1991}. The original plot is limited to wavelengths below 4000{\AA}; the spectral radiance is then extended to 6000{\AA} by fitting the 3000-4000{\AA} tail to a blackbody radiation curve. Since the radiance contributions of Hydrogen, Oxygen, and Nitrogen are treated separately, their emission features are suppressed here to avoid double-counting. Figure \ref{fig:earth_disk_spectrum} shows the final Earth disk continuum spectrum $\ell_\text{disk}(\lambda)$ [phot/s/cm$^2$/{\AA}]. The spectral radiance is assumed to be constant over all disk pixels, which neglects realistic spatial features such as limb darkening. However, the simplification is acceptable because Earth's disk acts primarily as a source of broad-spectrum contamination for the nearby science pixels. As the instrument's optical response blurs high-frequency spatial details, the exact distribution of radiance across the disk is less critical than the total flux available to bleed into the region of science interest.

\begin{figure}[htbp]
    \begin{subfigure}[t]{0.48\textwidth}
      \includegraphics[width=\textwidth]{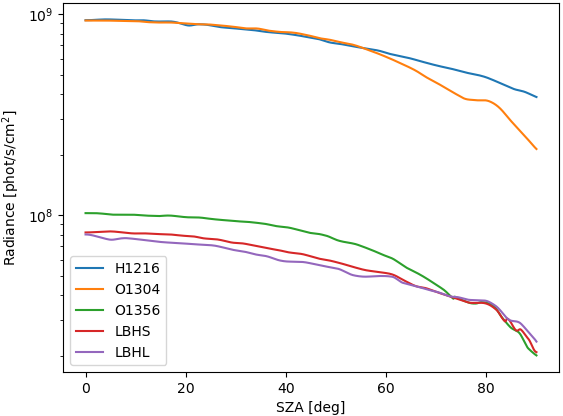}
      \caption{Total radiance as a function of solar zenith angle (SZA), from GUVI data on day 71 of 2009, for Hydrogen and Oxygen line emissions and Nitrogen emission bands.}
      \label{fig:guvi_radiance_vs_sza}
    \end{subfigure}
    \hfill
    \begin{subfigure}[t]{0.48\textwidth}
      \includegraphics[width=\textwidth]{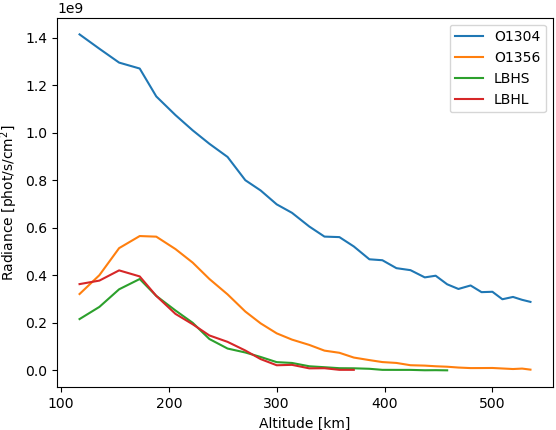}
      \caption{Total radiance as a function of altitude [km], from GUVI data, for Oxygen line emissions and Nitrogen emission bands.}
      \label{fig:guvi_radiance_vs_altitude}
    \end{subfigure}
    \caption{}
\end{figure}

The OOB spectral radiance contribution from Oxygen and Nitrogen in Earth's disk is modeled by analyzing data from GUVI, since physics-based models, such as the Atmospheric Ultraviolet Radiance Integrated Code (AURIC) \cite{strickland1999auric}, are computationally expensive. From 2001 to 2007, GUVI observed the upper atmospheric layers of Earth in the far-ultraviolet (FUV) wavelength range from an altitude of 625km and captured emission line/band radiances in the lower atmosphere as a function of solar zenith angle (SZA, denoted $\theta_{\text{sza}}$), where SZA is defined as the angle between the direct rays of the Sun and the vertical direction (i.e., directly overhead) at a location on Earth. However, the magnitudes of the measurements from GUVI depend on the season, year, and even time of day. To maintain generality, the OOB scene model uses GUVI data from a specific season and day (day 71 of 2009) to derive empirical scaling factors $c_{\text{disk}, m}(\theta_{\text{sza}}[i, j])$ that relate in-band Ly-$\alpha$ radiance to each OOB radiance $m$ as a function of SZA $\theta_{\text{sza}}[i, j]$ for pixel indices $i,j$. The SZA $\theta_{\text{sza}}[i, j]$ can be calculated using the GCI's viewing geometry and varies between $0^{\circ}$ to $90^{\circ}$ when $r_{ij} < 1$. The specific season and day chosen shows typical emission ratios on the dayside and their correspondence to SZA \cite{far_ultraviolet_instrument_technology} (see Figure \ref{fig:guvi_radiance_vs_sza}). Thus, the OOB spectral radiance contribution for the emission line/band $m$ in pixel $i,j$ in Earth's disk, in units of [phot/s/cm$^2$], is given as \begin{equation*}
    i_{\text{Ly-$\alpha$}}(i,j) c_{\text{disk},m}(\theta_{\text{sza}}[i, j])g_m(\lambda)
\end{equation*}

The complete scene model of OOB radiance in GCI pixels spanning Earth's disk is now complete. GUVI also measured the vertical structure of the four OOB spectral radiances of interest as a function of altitude above the Earth's surface, as shown in Figure \ref{fig:guvi_radiance_vs_altitude}. The OOB scene model normalizes GUVI's measured radiances against the lowest-altitude GUVI datapoint (100km) to obtain a scaling factor $c_{\text{low},m}(r_{ij})$. The final OOB radiance for emission line/band $m$ in the altitude range where GUVI data is present, i.e. up to 350km above the Earth's surface, or 1.055Re, is found by multiplying the scaling factor $c_{\text{low},m}(r_{ij})$ with the OOB radiance from the nearest Earth-disk pixel (smallest Euclidean distance in the image plane) to yield, in [phot/s/cm$^2$],
\begin{equation*}
    c_{\text{low},m}(r_{ij}) i_{\text{Ly-$\alpha$}}(p_{ij}, q_{ij}) c_{\text{disk},m}(\theta_{\text{sza}}[p_{ij}, q_{ij}])g_m(\lambda)
\end{equation*}

Finally, an exponential decay of the form $\beta[p_{ij}, q_{ij}] e^{-\alpha[p_{ij}, q_{ij}] \times (r_{ij} - 1)}$ is fit to GUVI data from altitudes 250-350km in order to extrapolate the OOB radiance at higher altitudes (up to 3150km). An independent exponential is fitted for each set of pixels $\boldsymbol{J}$ whose closest pixel in the disk is the same $p_{ij}, q_{ij}$. Above the altitude of 3150km OOB radiance is negligible compared to in-band radiance, so all OOB radiances are set to zero in this region.

For the Oxygen emission lines, it is convenient to rewrite Oxygen spectral radiance $\ell_{O1304}(\lambda, i, j)$ [phot/s/cm$^2$/{\AA}] in terms of Oxygen total radiance $i_{O1304}(i, j)$ [phot/s/cm$^2$], yielding $\ell_{O1304}(\lambda, i, j) = i_{O1304}(i, j) g_{O1304}(\lambda)$ and $\ell_{O1356}(\lambda, i, j) = i_{O1356}(i, j) g_{O1356}(\lambda)$ across all regions. The OOB scene model spectral radiance can now be written using two terms: a summation of $\delta$ functions (the two Oxygen line emissions, indexed by $m$) and a second term containing all emission bands, indexed by $n$. The OOB spectral radiance $\ell_{\text{OOB}}(\lambda, i, j)$ [phot/s/cm$^2$/\AA] can thus be expressed as:
\begin{multline}
    \label{eq:oob_scene_model_eq}
    \ell_{\text{OOB}}(\lambda, i, j) = \\
    \begin{cases}
        \left(\sum_m i_{m}(i, j) \delta(\lambda - \lambda_{m}) \right) + i_{\text{Ly-$\alpha$}}(i, j) \sum_n c_{\text{disk},n}(\theta_{\text{SZA}}[i,j])g_n(\lambda) + \ell_\text{disk}(\lambda) & \text{$r_{ij} < 1$} \\
        \left(\sum_m i_{m}(i, j) \delta(\lambda - \lambda_{m}) \right) + i_{\text{Ly-$\alpha$}}(p_{ij}, q_{ij})\sum_n c_{\text{low},n}(r_{ij})c_{\text{disk},n}(\theta_{\text{SZA}}[p_{ij}, q_{ij}])g_n(\lambda) & \text{$1 < r_{ij} < 1.055$} \\
        \left(\sum_m i_{m}(i,j) \delta(\lambda - \lambda_{m}) \right) + \sum_n \beta[p_{ij}, q_{ij}] \exp(-\alpha[p_{ij}, q_{ij}]\times (r_{ij} - 1))g_n(\lambda) & \text{$1.055 < r_{ij} < 1.5$} \\
        0 & \text{$1.5 < r_{ij}$} \\
    \end{cases}
\end{multline}

\subsection{Celestial Scene Model}

The stars, the Moon, and the outer planets also contribute some spectral radiance if they appear in the camera Field-Of-View (FOV). Stellar spectra are discussed in Zhang et al. (2026) \cite{Zhang26c}; the scene model below assumes that the spectra are already known. Stars, indexed by $k$, are modeled as point sources that are projected onto pixel $i_k, j_k$.
\begin{equation*}
    \ell_{\text{star}}(\lambda, i, j) = \sum_k \ell_k(\lambda) \delta(i - i_k) \delta(j - j_k)
\end{equation*}

The lunar spectrum is obtained by multiplying the SUSIM solar spectrum \cite{SUSIM_mission_paper} by 0.06 (which is the mean lunar albedo \cite{MoonFlynn98, MoonLiu18}). The SUSIM solar spectrum is limited to wavelengths below 3200{\AA}; it is then extended to 6000{\AA} by scaling the Earth spectrum at these wavelengths so that the mean spectral radiance at the wavelengths between 3000{\AA} and 3200{\AA} matches for both spectra. See Figure \ref{fig:lunar_spectrum} for the final lunar spectrum. The lunar mask, denoted $m_{\text{Moon}}(i, j)$, delineates the specific set of pixels onto which the Moon is projected in the camera frame.
\begin{equation*}
    \ell_{\text{Moon}}(\lambda, i, j) = \ell_{\text{Moon}}(\lambda) m_{\text{Moon}}(i, j)
\end{equation*}

\begin{figure}[htbp]
    \begin{subfigure}[t]{0.48\textwidth}
      \includegraphics[width=\textwidth]{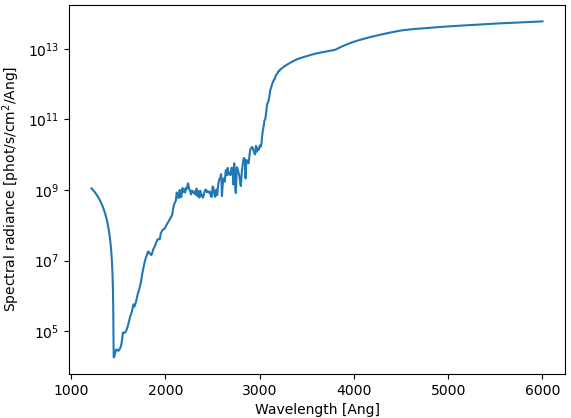}
      \caption{Mars' spectral radiance}
      \label{fig:mars_spectrum}
    \end{subfigure}
    \hfill
    \begin{subfigure}[t]{0.48\textwidth}
      \includegraphics[width=\textwidth]{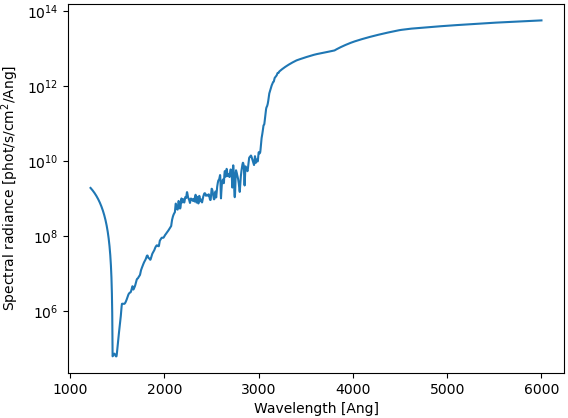}
      \caption{Jupiter's spectral radiance}
      \label{fig:jupiter_spectrum}
    \end{subfigure}
    \caption{Spectral radiance for Mars and Jupiter. The remaining outer planets exhibit spectral radiance profiles analogous to Jupiter's, varying only in absolute magnitude.}
    \label{fig:planet_spectrum}
\end{figure}

The five outer planets (Mars, Jupiter, Saturn, Uranus, and Neptune) will also occasionally appear in the FOV and contribute additional spectral radiance. The scene model only requires a first-order estimate of each planet's brightness as seen by the instrument. Since the planets appear as contamination sources rather than calibration targets, exact photometry is unnecessary; the brightness model serves primarily to ensure that the simulated signal dynamic range is representative of on-orbit conditions. Thus, the scene model prioritizes simplicity over accuracy when modeling the five outer planets. For all five planets, the solar spectrum \cite{SUSIM_mission_paper} is extended out to 6000{\AA} using the same process as described before with the lunar spectra. Next, the spectrum is multiplied scaled by $0.67$. This factor accounts for the Lambertian limb darkening of the planetary sphere, correcting the uniform disk model to match the spatially integrated flux of a diffuse reflector \cite{hapke2012theoryalbedo} to account for the limb darkening effect. Finally, the spectrum is multiplied by each planet's albedo. The gas giants all use the same albedo curve, with data in the EUV (extreme ultra-violet, or wavelengths $<1300$ {\AA}) from Morrissey et al. (1995) \cite{morrissey1995euvalbedojupiter} and data in the FUV (far ultra-violet, or wavelengths $>1300${\AA}) from Clarke et al. (1982) \cite{clarke1982fuvalbedojupiter}. Mars exhibits a different albedo curve, with data in the EUV from McCord et al. (1971) \cite{mccord1971euvalbedomars} and data in the FUV from Clarke et al. (2014) \cite{clarke2014fuvalbedomars}. See Figure \ref{fig:planet_spectrum} for the final spectra. Each planet also has an accompanying mask over image pixels, denoted $m_{\text{planet}}(i, j)$, which delineates the specific set of pixels onto which the planet is projected in the camera frame.
\begin{equation*}
    \ell_{\text{planets}}(\lambda, i, j) = \sum_{\text{planet}} \ell_{\text{planet}}(\lambda) m_{\text{planet}}(i, j)
\end{equation*}

\subsection{Total UV Photon Scene Model}

The final spectral radiance $\ell(\lambda, i, j)$ [phot/s/cm$^2$/{\AA}] is the sum of each individual contribution. In Equation \ref{eq:all_photon_sources} below, the stellar contribution $\ell_{\text{stars}}(\lambda, i, j)$ requires special treatment: unlike all the other sources, which are diffuse, stars are sub-pixel point sources. Therefore, the stellar term is scaled by $\frac{4\pi}{\Omega}$ (where $\Omega$ is the pixel solid angle) to convert the point-source flux into an equivalent pixel-averaged radiance, ensuring consistency with the diffuse components.
\begin{equation}
    \label{eq:all_photon_sources}
    \ell(\lambda, i, j) = \ell_{\text{exo}}(\lambda, i, j) + \ell_{\text{IPH}}(\lambda, i, j) + \ell_{\text{OOB}}(\lambda, i, j) + \frac{4\pi}{\Omega}\ell_{\text{stars}}(\lambda, i, j) + \ell_{\text{Moon}}(\lambda, i, j) + \ell_{\text{planets}}(\lambda,i,j)
\end{equation}

The summation in Equation \ref{eq:all_photon_sources} assumes that the Earth is within the FOV. Images that do not have Earth within the FOV (such as calibration images) are referred to as `off-nadir' images. Simulation of these images requires setting $\ell_{\text{exo}}(\lambda, i, j) = \ell_{\text{OOB}}(\lambda, i, j) = 0$ for all $\lambda, i, j$.

\section{Instrument Model}
\label{sec:instrument_model}

This section describes the Carruthers GeoCoronal Imager (GCI) instrument model that converts incident spectral radiance (modeled as described in the previous section) into the measured images with pixel signals in instrument units of [Digital Numbers], or [DN]. The scene model described in the previous section was assumed to be time-invariant during a given image integration; for generality, the instrument model below introduces a time dependence $t$ in the spectral radiance. Thus, in this section, spectral radiance is denoted as $\ell(\lambda, i, j, t)$ [phot/s/cm$^2$/{\AA}].

Deriving the observed spectral radiance, $\ell_{\text{obs}}(\lambda, i, j, t)$ [phot/s/cm$^2$/{\AA}], from the incident field, $\ell(\lambda, i, j, t)$ [phot/s/cm$^2$/{\AA}], requires applying a sequence of operators that simulate the instrument's physical response. The process begins with an optical distortion operator, $d(\cdot)$, which maps the incident radiance onto the detector's coordinate grid, followed by convolution with the system's Point Spread Function (PSF), $h(i, j)$. The signal is then integrated over the pixel solid angle fraction $\Omega/4\pi$ (assuming pixel spatial uniformity) and modulated by the optical flat-field term, $f_{\text{opt}}(i, j)$, which accounts for spatial variations in transmission efficiency such as vignetting. Collecting these terms, the effective spectral radiance $\ell_{\text{obs}}(\lambda, i, j, t)$ [phot/s/cm$^2$/{\AA}] reaching the detector plane is given by
\begin{equation*}
    \ell_{\text{obs}}(\lambda, i, j, t) = f_{\text{opt}}(i, j) \frac{\Omega}{4\pi} \left[d(\ell(\lambda, i, j, t)) * h(i,j)\right]
\end{equation*}

The GCI is equipped with two channels, which each have independent open system optical efficiencies as a function of wavelength, denoted $\varepsilon(\lambda)$, with units of [photoelectron events/incident photon], or for short, [events/photon]. The open system optical efficiency depends on both the wavelength-dependent mirror reflectivities for mirror $m$, denoted $\rho_m(\lambda)$, and the photocathode quantum efficiency (QE), denoted $\eta(\lambda)$ [events/photon]. The two channels have a different number of mirrors, denoted $n_{\text{mir}}$. The formula for optical efficiency is as follows:
\begin{equation*}
    \varepsilon(\lambda) = \eta(\lambda)\prod_{m=1}^{n_{\text{mir}}}\rho_m(\lambda)
\end{equation*}

All photons that reach the detector plane, over aperture area (denoted by $a$), pass through these optics. Both channels are also equipped with independent $6$-position filter wheels to provide further UV spectral targeting options, as described in Sirk et al. (2026) \cite{sirk26a}. The wavelength-dependent filter transmissivity for filter $f$ is denoted by $\tau_f(\lambda)$. Therefore, the mean number of detected photons at time $t$ per pixel $i, j$ for filter $f$, in units of [photoelectron events/s], is
\begin{equation}
    \label{eq:ephoton_def}
    e_{\text{photon}, f}(i, j, t) = f_{\text{opt}}(i, j) \frac{a\Omega}{4\pi} \int_0^{\infty} \varepsilon(\lambda)\tau_f(\lambda)(d(\ell(\lambda, i, j, t)) * h(i, j))d\lambda
\end{equation}

The number of events per image frame $k$ can be well modeled as a Poisson process $E_{\text{photon}, f}(i, j, k)$ [events/frame] \cite{bertolotti1974photonpoisson} with mean $e_{\text{photon}, f}(i, j, k)$ equal to the time integral of $e_{\text{photon}, f}(i, j, t)$ over the exposure window (from time $t_k$ to time $t_k + t_{\text{frame}}$, where $t_{\text{frame}}$ denotes frame integration time).
\begin{equation}
    \label{eq:e_photon_framerate_definition}
    E_{\text{photon}, f}(i, j, k) \sim \text{Pois}(e_{\text{photon}, f}(i, j, k)), \ \ \ \ \ e_{\text{photon}, f}(i, j, k) = \int_{t_k}^{t_k + t_{\text{frame}}} e_{\text{photon}, f}(i, j, t) dt
\end{equation}
Subsequently, the filter subscript $f$ is dropped to reduce notational clutter, although it is implied in the remainder of this section.

In addition to photon events, the instrument is subject to a non-photon background caused by solar energetic particle radiation striking the Micro-Channel Plate (MCP), which generates a time-variable event rate, $e_{\text{rad}}(t)$ [events/s]. The instrument model assumes no spatial dependence on pixel coordinates $(i, j)$, which is justified because the stochastic particle impacts effectively average to a uniform background level over the integration time of the stacked image sequence. This source of photoelectron events is referred to as MCP radiation. For each image frame, the total number of radiation events is modeled as a Poisson process, $E_{\text{rad}}(k)$ [events/frame], where the expected value $e_{\text{rad}}(k)$ [events/frame] is the time-integral of $e_{\text{rad}}(t)$ over the exposure window (from $t_k$ to time $t_k + t_{\text{frame}}$):
\begin{equation*}
    E_{\text{rad}}(k) \sim \text{Pois}(e_{\text{rad}}(k)), \ \ \ \ \ e_{\text{rad}}(k) = \int_{t_k}^{t_k + t_{\text{frame}}} e_{\text{rad}}(t) dt
\end{equation*}
On-orbit, the mean depends on the environment of the spacecraft. Specific details on the time-variability and the typical values of MCP radiation mean can be found in Zhang et al. (2026) \cite{Zhang26a}.

The total number of photoelectron events per image frame is the sum of the individual sources of photoelectron events per image frame, which is a random variable given as
\begin{equation}
    \label{eq:e_ijk_equation}
    E(i, j, k) = E_{\text{photon}}(i, j, k) + E_{\text{rad}}(k)
\end{equation}

\begin{figure}[htbp]
    \begin{subfigure}[t]{0.48\textwidth}
      \includegraphics[width=\textwidth]{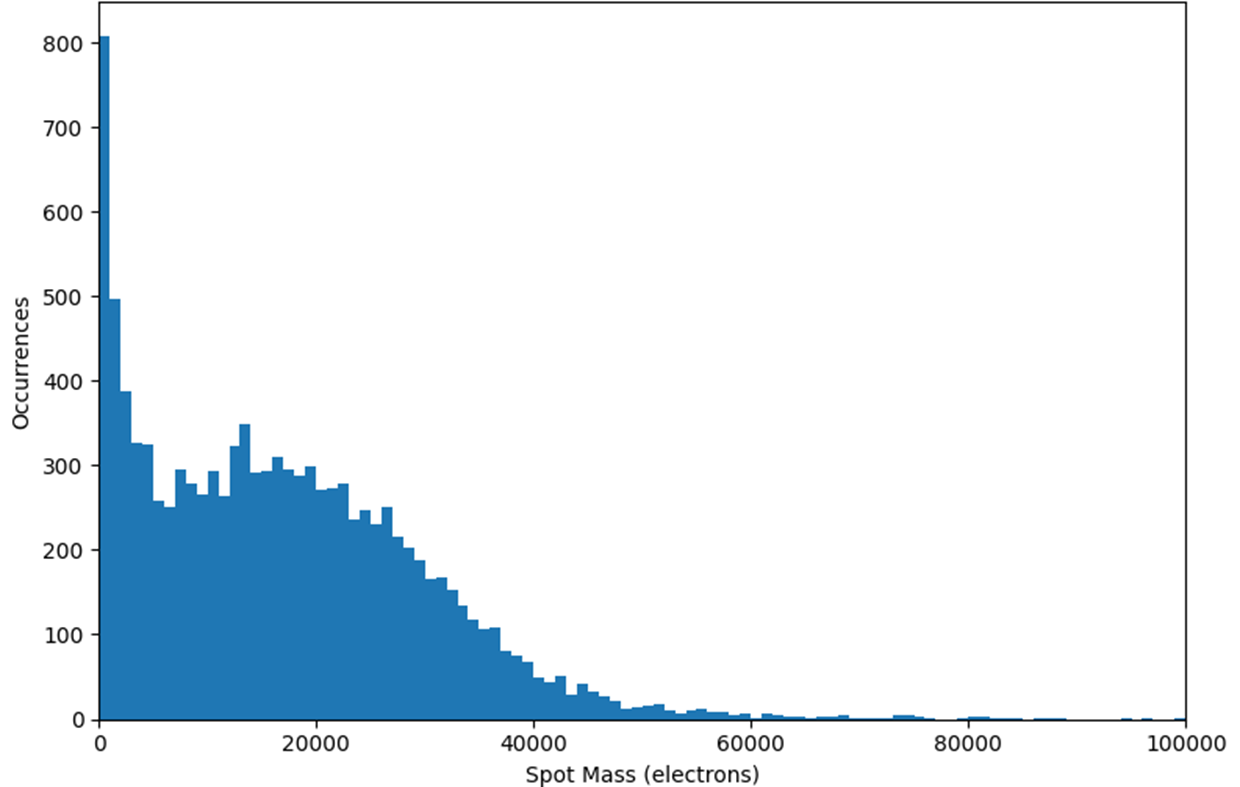}
      \caption{}
    \end{subfigure}
    \hfill
    \begin{subfigure}[t]{0.48\textwidth}
      \includegraphics[width=\textwidth]{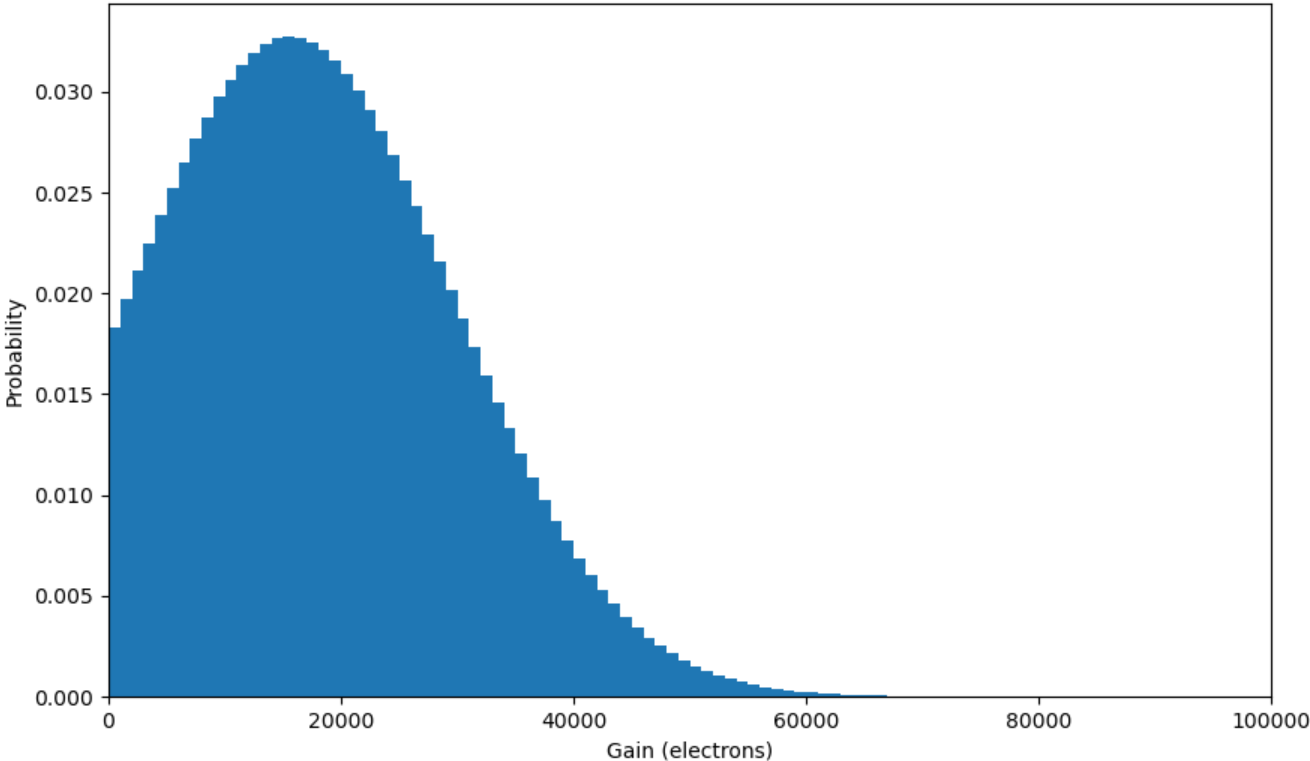}
      \caption{}
    \end{subfigure}
    \caption{Measured (left) and modeled (right) Micro-Channel Plate (MCP) Pulse Height Distributions (PHD). The measured distribution exhibits a clear modal gain centered at approximately 19,000 electrons. The sharp peak near zero represents the electronic noise pedestal, which was removed when obtaining the final modeled distribution. The measured PHD was obtained during pre-launch laboratory calibrations. \label{fig:mcp_gain_dist}}
\end{figure}

Any photoelectron or radiation ``event'' incident on the MCP undergoes gain amplification via an electron avalanche process. When a primary electron strikes the channel walls, it induces the release of multiple secondary electrons. These secondary electrons are accelerated by the high voltage potential down the channel, striking the walls repeatedly and releasing further electrons at each stage. The resulting multiplicative cascade amplifies a single input event into a measurable charge cloud of $\sim10^5$ electrons at the anode. These electrons excite a phosphor screen, which releases visible photons that accumulate in the $2080 \times 2048$ pixels of a silicon-based Active Pixel Sensor (APS) detector, resulting in electron counts. The number of electron counts per photoelectron event, denoted $G_{\text{mcp}}$ [counts/event], is controlled by the MCP voltage setting and does not follow any standard distribution (see Figure \ref{fig:mcp_gain_dist}).
Since $G_{\text{mcp}}$ is an empirical distribution with finite support, its statistical moments are well-defined and finite. The model assumes that the amplification of each photoelectron event is statistically independent. However, due to the MCP manufacturing process, the amplification efficiency is not spatially uniform across the detector face. Non-uniformities arise from stacking faults and channel distortions at the boundaries of the fused multifiber bundles, which locally alter the electron impact geometry and secondary emission efficiency. These pixel-to-pixel gain variations usually manifest as hexagonal patterns and is referred as the MCP flat-field, $f_{\text{mcp}}(i, j)$, which is normalized to have a spatial mean of unity. Consequently, the total electron counts per image frame $C_{\text{events}}(i, j, k)$ [counts/frame] due to photoelectron events is given by:
\begin{equation}
    \label{eq:countrate_from_events}
    C_{\text{events}}(i, j, k) = f_{\text{mcp}}(i, j) \sum_{n = 1}^{E(i, j, k)} G_{\text{mcp}, n}
\end{equation}
Here, each $G_{\text{mcp}, n}$ is assumed to be i.i.d.

The thermal electron current due to the thermal motion of the electrons within each channel manifests as additional electron counts per second, or ``dark current''. The mean number of electron counts per second as a result of dark current, denoted $c_{\text{dark}}(t)$ [counts/s], accumulates linearly with respect to time and is correlated with the temperature of the system. The number of electron counts per image frame $C_{\text{dark}}(k)$ [counts/frame] is modeled as a Poisson process with mean $c_{\text{dark}}(k)$ [counts/frame] equal to the integration of $c_{\text{dark}}(t)$ from time $t_k$ to time $t_k + t_{\text{frame}}$:
\begin{equation*}
    C_{\text{dark}}(k) \sim \text{Pois}(c_{\text{dark}}(k)), \ \ \ \ \ c_{\text{dark}}(k) = \int_{t_k}^{t_k + t_{\text{frame}}} c_{\text{dark}}(t) dt
\end{equation*}

Non-photon solar energetic particle radiation that are incident on the APS can create multiple electron-hole pairs in the silicon (Si), which has the net effect of creating additional electron counts per second, denoted $c_{\text{rad}}(t)$ [counts/s]. This source of instrument background signal is referred to as APS radiation. The contribution per image frame is modeled as a Poisson process $C_{\text{rad}}(k)$ [counts/frame] with mean $c_{\text{rad}}(k)$ [counts/frame] equal to the integration of $c_{\text{rad}}(t)$ from time $t_k$ to time $t_k + t_{\text{frame}}$:
\begin{equation*}
    C_{\text{rad}}(k) \sim \text{Pois}(c_{\text{rad}}(k)), \ \ \ \ \ c_{\text{rad}}(k) = \int_{t_k}^{t_k + t_{\text{frame}}} c_{\text{rad}}(t) dt
\end{equation*}
The mean $c_{\text{rad}}(k)$ depends on the environment of the spacecraft. Note that an energetic particle can either ``create'' additional electron counts or additional photoelectron events, but not both. Any particle radiation incident on the MCP will undergo the intensifier gain and is modeled as additional photoelectron events, while the particle radiation directly incident on the APS does not undergo intensifier gain and is modeled as electron counts. The energetic particle flux in the spacecraft environment is the same in both cases, but the radiation signal detected varies with the material properties of the MCP and the APS sensor.

Anomalous pixels, or pixels with abnormally high or abnormally low electron countrates not explained by any of the above sources, can also appear due to aberrant leakage current on some component or fabrication variation. The number of electron counts in an image frame $k$ per pixel $i, j$ is modeled as a Poisson-distributed random variable $C_{\text{anom}}(i, j, k)$ [counts/frame] with mean $c_{\text{anom}}(i, j, k)$ [counts/frame].

The APS readout architecture is divided into two independent chains processing the top and bottom halves of the sensor, respectively. Each chain consists of a dedicated Field Programmable Gate Array (FPGA) interfaced with an Analog-to-Digital Converter (ADC) - one for the top half of the sensor and one for the bottom half, each of which result in a half-image of dimensions $1040 \times 2048$. Within each half-image, there are $8$ electrically dark rows, $8$ optically dark rows, and $1024$ regular rows. Electrically dark rows have their photodiodes disabled, which effectively sets all sources of counts $C$ to zero for all pixels in these rows for all frames. Optically dark rows have a metal mask over their photodiodes, so they cannot see light (i.e. $C_{\text{events}}(i, j, k) = 0$). However, other sources of instrument background with units of [counts] are still valid. Finally, pixels located in the corners of the image plane exhibit effectively zero gain due to the optical coupling between the intensifier and the detector. The fiberoptic taper, which transfers the optical signal from the phosphor screen to the sensor surface, acts as a circular aperture. Because the taper's output diameter is smaller than the diagonal of the rectangular detector array, the corner pixels lie outside the bonded fiber interface. Consequently, these pixels are optically isolated from the phosphor signal ($C_{\text{events}} = 0$), though they remain electrically active and continue to accumulate non-photon instrument background (e.g., dark current, read noise).  Pixels with gain amplification are said to be within the field-of-view (FOV) for the appropriate channel. Pixels without gain amplification are said to be outside the FOV.

In summary, the number of electron counts $C(i, j, k)$ per frame $k$ and per pixel $i, j$, in units of [counts/frame], is given as
\begin{equation}
    \label{eq:counts_full_model}
    C(i, j, k) = \begin{cases}
        C_{\text{rad}}(k) + C_{\text{dark}}(k) + C_{\text{anom}}(i, j, k) + C_{\text{events}}(i, j, k) & \text{$i,j$ in FOV} \\
        C_{\text{rad}}(k) + C_{\text{dark}}(k) + C_{\text{anom}}(i, j, k) & \text{$i,j$ outside of FOV} \\
        C_{\text{rad}}(k) + C_{\text{dark}}(k) + C_{\text{anom}}(i, j, k) & \text{$i,j$ in optically dark row} \\
        0 & \text{$i,j$ in electrically dark row} \\
    \end{cases} 
\end{equation}

The ADC carries a constant gain factor as it converts electron counts to Digital Numbers (DN), denoted $g_{\text{adc}}$ [DN/count], where the factor $g_{\text{adc}}$ represents the quantization step size of the digitization process. The readout process is accompanied by read noise, which is modeled as a Gaussian distribution with zero mean $R$. Finally, a voltage bias $b(i, j, k)$ [DN/frame] is applied before readout. The voltage is assumed to be a deterministic electronic offset and is thus modeled as a constant.

Therefore, the signal $S(i, j, k)$ per frame $k$ and per APS pixel $i, j$, in units of [DN/frame], is given as
\begin{equation}
    \label{eq:DN_source_eq}
    S(i, j, k) = R(i, j, k) + b(i, j, k) + g_{\text{adc}}C(i, j, k)
\end{equation}

After read-out, the images are binned by $n_{\text{bin}} \times n_{\text{bin}}$ pixels, where $n_{\text{bin}} = 2$ for the NFI channel and $n_{\text{bin}} = 4$ for the WFI channel to decrease the size of each image so that the image data can meet the Carruthers mission's telemetry budget constraints. Therefore, the final NFI images to have dimensions of $1040 \times 1024$, while the final WFI images have dimensions of $520 \times 512$.

The signal $S(p, q, k)$ per frame $k$ and per binned channel pixel $p, q$, in units of [DN/frame], is given as
\begin{equation*}
    S(p, q, k) = \sum_{i=pn_{\text{bin}}}^ {(p+1)n_{\text{bin}} - 1} \sum_{j=qn_{\text{bin}}}^{(q+1)n_{\text{bin}} - 1} S(i, j, k)
\end{equation*}

Image frames are stacked onboard before being telemetered back to Earth via the Instrument Control Package (ICP). The number of frames stacked, denoted $n_{\text{frames}}$, is a function of required exposure time $t_{\text{int}}$ [seconds] and the frame integration time [seconds]. In the baseline operations plan, individual frame integration time, or $t_{\text{frame}}$, is $0.125$ seconds. Most telemetered images will have an integration time between $1$ and $60$ minutes.
\begin{equation*}
    n_{\text{frame}} = \frac{t_{\text{int}}}{t_{\text{frame}}}
\end{equation*}

Therefore, the total measured signal in the final image $S(p, q)$ per binned channel pixel $p, q$, in units of [DN], is given as
\begin{equation}
    \label{eq:final_DN_equation}
    S(p, q) = \sum_{k = 1}^{n_{\text{frame}}} S(p, q, k)
\end{equation}

\section{Numerical Simulation}

Numerical simulation of images is an important tool to validate the performance of the calibration pipeline. The numerical simulation was implemented using Python and follows the description of both models in the previous two sections. However, Equation \ref{eq:countrate_from_events} presents a computational issue, as it requires a large number of independent draws from a non-standard random distribution $G_{\text{mcp}}$. However, the Central Limit Theorem (CLT) provides a robust justification for approximating the resulting distribution as Gaussian since the total signal is derived by aggregating discrete photoelectron events over $n_{\text{frames}}$ independent frames. As demonstrated in Appendix \ref{sec:clt_approx_validation}, the approximation remains valid for accumulated signals exceeding a threshold of $1784$ [events]. Given the standard exposure duration of $30$ minutes, the derived threshold corresponds to a mean event rate of approximately $1$ [event/s]. Primary signal sources, including the low-altitude exosphere, the OOB background, and stellar point sources, are reliably above this mean event rate. In regimes where the source flux falls below this limit (i.e., dim sources), the failure of the Gaussian approximation is negligible since the signal in these cases is dominated by the instrument background rather than the source variance.

Further simplifying approximations are also justified. The summation of multiple independent Poisson random variables in Equation \ref{eq:counts_full_model} is itself a Poisson random variable, which can also be modeled as an appropriate Gaussian random variable since the two distributions converge in distribution as the mean of the Poisson random variable grows larger (see Appendix \ref{sec:poisson_to_gaussian}). Therefore, the total contribution, in [counts], to the final image, excluding countrate contributions from photoelectron events, can be approximated as a Gaussian random variable, as the smallest possible mean in Equation \ref{eq:counts_full_model} can be lower-bounded by assuming that $C_{\text{rad}}(k) = C_{\text{anom}}(i,j,k) = 0$, then setting $C_{\text{dark}}(k)$ to its smallest baseline value at $20$ [counts] per binned pixel per second (measured during pre-launch laboratory calibrations). Thus, in a typical $30$-minute integration of a science image, a binned pixel will collect a minimum of $36,000$ [counts] on average due to dark current alone, which is well over the $5,191$ [counts] needed for these Poisson RVs to be well-approximated by a Gaussian RV as shown in Appendix \ref{sec:poisson_to_gaussian}. Furthermore, a $5$-minute dark exposure is sufficiently long to justify this approximation since a binned pixel will collect a minimum of $6,000$ [counts] on average.

Finally, since the summation of independent Gaussian random variables results in another Gaussian random variable, $C(i, j, k)$ in Equation \ref{eq:counts_full_model} can be approximated as a Gaussian distribution, implying that $S(p, q)$ from Equation \ref{eq:final_DN_equation} can also be approximated as Gaussian, which reduces the number of random draws required for numerical simulation of the GCI instrument response to simply one random draw per binned pixel (instead of one random draw per photon). The mean and variance of $S(p, q)$ are derived analytically in Appendix \ref{sec:raw_im_derivs}.

Finally, the numerical image simulator was designed with the flexibility to enable/disable individual sources of events, counts, and DN to allow for faster synthetic image generation.

\section{Conclusion}

This paper presented a comprehensive forward model and numerical simulation framework for the GeoCoronal Imager (GCI) aboard the Carruthers Geocorona Observatory. To synthesize realistic raw images, the scene model quantified individual sources of incident spectral radiance fields mapped from the Carruthers viewing geometry at the Earth-Sun L1 Lagrange point. Spectral radiance fields included the target exospheric Lyman-$\alpha$ emission, alongside the non-terrestrial InterPlanetary Hydrogen (IPH) Ly-$\alpha$ background, terrestrial Out-Of-Band (OOB) emissions, and celestial sources such as stars, the Moon, and the outer planets.

The instrument model detailed the physical and electronic transformations that map this incident spectral radiance field to the instrument's raw output in Digital Numbers (DN). This included optical distortions, point spread functions, filter and mirror efficiencies, and a stochastic accounting of instrument noise sources. Key modeled variables included Micro-Channel Plate (MCP) amplification and radiation, dark current, and Active Pixel Sensor (APS) radiation events. By implementing these models in Python and leveraging the Central Limit Theorem to approximate Poissonian and computationally expensive non-standard distributions as Gaussian, the numerical simulator generates high-fidelity synthetic images efficiently. 

Ultimately, this numerical simulation serves as a foundational and computationally efficient tool for the Carruthers mission. The synthetic images generated by this framework will be heavily utilized in other papers in this special issue to conduct Monte Carlo validations of various calibration algorithms, hydrogen density retrieval processes, and the full data processing pipeline. Due to mission restrictions, the numerical simulation code is currently proprietary; please contact the first author with inquiries.

\addcontentsline{toc}{section}{References}
\printbibliography 

\appendix

\section{Raw Image Statistics Derivations}
\label{sec:raw_im_derivs}

\subsection{Convergence of High-Mean Poisson Random Variables to Gaussian Random Variables}
\label{sec:poisson_to_gaussian}

It is known that the summation of two independent Poisson Random Variables (RVs) with means $x$ and $y$ is itself a Poisson RV with mean $x+y$. The proof can be found in an undergraduate probability textbook. Therefore, Poisson RVs with a large mean $x$ can be decomposed into a summation of $x$ independent Poisson RVs each with mean $1$. Summations of i.i.d. random variables converge in distribution to a Gaussian RV by the Central Limit Theorem, which in this case will have mean equal to variance equal to $x$. Assuming that the discretization of the result is not important (as is the case for applications of this proof in this paper), this section gives a statistical argument for what mean $x$ is considered large enough for the Poisson RV to be replaced by a Gaussian RV.

The Berry-Esseen bound \cite{esseen1942berryesseenbound} for the sum of $n$ i.i.d. Poisson RVs $X_i$, $1 \leq i \leq n$, with mean $1$ centered and standardized with cumulative distribution function (CDF) $F_n(\cdot)$ is given as
\begin{equation*}
    \sup_x\left\vert F_n(x) - \Phi(x)\right\vert \leq \frac{C\EX[|X-\EX[X]|^3]}{\sqrt{n}}
\end{equation*}
For a Poisson RV with mean $1$, the third absolute moment $\EX[|X-\EX[X]|^3]$ can be calculated to be about 1.7357588823 \cite{johnson2005univariatedistributionsbook}. This satisfies the bound $\rho \geq 1.286\sigma^3 = 1.286$, so we can apply the tighter bound found by Shevtsova (2011) \cite{shevtsova2011berryesseenconstant} (again, applying $\sigma = 1$):
\begin{equation*}
    \sup_x\left\vert F_n(x) - \Phi(x)\right\vert \leq \frac{0.3328(\rho + 0.429)}{\sqrt{n}} = \frac{0.720431756029}{\sqrt{n}}
\end{equation*}

A maximum CDF error $\varepsilon = \sup_x\left\vert F_n(x) - \Phi(x)\right\vert$ of $0.01$ represents a theoretical uncertainty of 1\%. The real-world instrumentation that applications of this approximation aim to model often has systematic uncertainties (such as thermal noise, transient artifacts, shot noise, etc.) that frequently exceed 1\%. Therefore, we can set the left-hand side of the above to $0.01$, which implies that we require Poisson distributions to have a mean $x \geq 5191$. Chasing a tighter bound would demand drastically higher counts without improving the fidelity of the model. Note that the usual applications of this bound in this paper are on Poisson distributions with means well into the $10^4$ range, which means there is plenty of margin.

\subsection{Mean Counts from Events}
\label{sec:mean_counts}

This section derives $\EX[C_{\text{events}}(i, j, k)]$, whose formula is given in Equation \ref{eq:countrate_from_events}.

\begin{equation*}
    \EX[C_{\text{events}}(i, j, k)] = \EX\left[f_{\text{mcp}}(i, j) \sum_{n = 1}^{E(i, j, k)} G_{\text{mcp}, n} \right]
\end{equation*}
Invoke Law of the Unconscious Statistician (LOTUS):
\begin{equation*}
    = f_{\text{mcp}}(i, j) \sum_x p(E(i, j, k) = x) \EX\left[\sum_{n = 1}^{E(i, j, k)} G_{\text{mcp}, n}\middle\vert E(i,j,k) = x\right]
\end{equation*}
\begin{equation*}
    = f_{\text{mcp}}(i, j) \sum_x p(E(i, j, k) = x) x \EX[G_{\text{mcp}}]
\end{equation*}
\begin{equation*}
    = f_{\text{mcp}}(i, j) \EX[G_{\text{mcp}}] \EX[E(i, j, k)]
\end{equation*}
\begin{equation*}
    = f_{\text{mcp}}(i, j) \EX[G_{\text{mcp}}] (e_{\text{photon}}(i,j,k) + e_{\text{rad}}(k))
\end{equation*}

\subsection{Variance of Counts from Events}
\label{sec:var_counts}

This section derives $\text{Var}[C_{\text{events}}(i, j, k)]$, whose formula is given in Equation \ref{eq:countrate_from_events}. Define the random variable $X = \sum_{n = 1}^{E(i, j, k)} G_{\text{mcp}, n}$.
\begin{equation*}
    \text{Var}[C_{\text{events}}(i, j, k)] = \text{Var}\left[f_{\text{mcp}}(i, j) \sum_{n = 1}^{E(i, j, k)} G_{\text{mcp}, n} \right] = f_{\text{mcp}}(i, j)^2\text{Var}\left[X\right]
\end{equation*}
\begin{equation*}
    = f_{\text{mcp}}(i, j)^2\left(\EX\left[X^2\right] - \EX\left[X\right]^2\right)
\end{equation*}
Invoke tower property:
\begin{equation*}
    = f_{\text{mcp}}(i, j)^2\left(\EX\left[\EX\left[X^2\middle\vert E(i,j,k)\right]\right] - \EX\left[\EX\left[X\middle \vert E(i,j,k)\right]\right]^2\right)
\end{equation*}
\begin{equation*}
    = f_{\text{mcp}}(i, j)^2\left(\EX\left[\text{Var}[X\middle\vert E(i,j,k)] + \EX[X\middle\vert E(i,j,k)]^2\right] - \EX\left[\EX\left[X\middle \vert E(i,j,k)\right]\right]^2\right)
\end{equation*}
\begin{equation*}
    = f_{\text{mcp}}(i, j)^2\left(\EX\left[\text{Var}[X \middle\vert E(i,j,k)]\right] + \EX\left[\EX[X\middle\vert E(i,j,k)]^2\right] - \EX\left[\EX\left[X\middle\vert E(i,j,k)\right]\right]^2\right)
\end{equation*}
\begin{equation*}
    = f_{\text{mcp}}(i, j)^2\left(\EX\left[\text{Var}[X\middle\vert E(i,j,k)] \right] + \text{Var}\left[\EX[X\middle\vert E(i,j,k)] \right]\right)
\end{equation*}
\begin{equation*}
    = f_{\text{mcp}}(i, j)^2\left(\EX\left[\text{Var}\left[\sum_{n = 1}^{E(i, j, k)} G_{\text{mcp}, n}\middle\vert E(i,j,k)\right] \right] + \text{Var}\left[\EX\left[\sum_{n = 1}^{E(i, j, k)} G_{\text{mcp}, n}\middle\vert E(i,j,k)\right] \right]\right)
\end{equation*}
Use the fact that each $G_{\text{mcp}, n}$ is i.i.d.:
\begin{equation*}
    = f_{\text{mcp}}(i, j)^2\left(\EX\left[\sum_{n = 1}^{E(i, j, k)} \text{Var}\left[G_{\text{mcp}, n}\middle\vert E(i,j,k)\right] \right] + \text{Var}\left[\sum_{n = 1}^{E(i, j, k)} \EX\left[G_{\text{mcp}, n}\middle\vert E(i,j,k)\right] \right]\right)
\end{equation*}
Since $G_{\text{mcp}, n}$ and $E(i,j,k)$ are also independent for all $n$, we can drop the condition in both terms. Finally, all $G_{\text{mcp}, n}$ are i.i.d., so we can drop the dependence on $n$.
\begin{equation*}
    = f_{\text{mcp}}(i, j)^2\left(\EX\left[\sum_{n = 1}^{E(i, j, k)} \text{Var}\left[G_{\text{mcp}}\right] \right] + \text{Var}\left[\sum_{n = 1}^{E(i, j, k)} \EX\left[G_{\text{mcp}}\right] \right]\right)
\end{equation*}
\begin{equation*}
    = f_{\text{mcp}}(i, j)^2\left(\EX\left[E(i, j, k)\right]\text{Var}\left[G_{\text{mcp}}\right]  + \text{Var}\left[E(i, j, k) \right]\EX\left[G_{\text{mcp}}\right] ^2\right)
\end{equation*}
Recall that $E(i,j,k)$ is Poisson-distributed with mean $e_{\text{photon}}(i,j,k) + e_{\text{rad}}(k)$ and mean equal to its variance.
\begin{equation*}
    = f_{\text{mcp}}(i, j)^2\left(e_{\text{photon}}(i,j,k) + e_{\text{rad}}(k)\right)\left(\text{Var}\left[G_{\text{mcp}}\right] + \EX\left[G_{\text{mcp}}\right]^2\right)
\end{equation*}
\begin{equation*}
    = f_{\text{mcp}}(i, j)^2\EX\left[G_{\text{mcp}}^2\right]\left(e_{\text{photon}}(i,j,k) + e_{\text{rad}}(k)\right)
\end{equation*}

\subsection{Validation of CLT Approximation}
\label{sec:clt_approx_validation}

Since the summation $X = \sum_{i = 1}^{N} G_{\text{mcp}, i}$ can be quite computationally intensive to evaluate by individually drawing each sample from $G_{\text{mcp}, i}$, the numerical image simulator invokes the Central Limit Theorem (CLT) to obtain a random instance of this summation. This section justifies the validity of this approach.

Korolev and Shevtsova (2012) \cite{korolev2012randomsumbound} found a bound on the largest absolute difference between the CDF of the sum of a random number $N$ of i.i.d. RVs compared to the Gaussian approximation, which is based on the Berry-Esseen bound \cite{esseen1942berryesseenbound}, where $N$ is Poisson-distributed with mean $z$. Applying this bound to the summation $X$ yields
\begin{equation*}
    \sup_x \left\vert P\left(\frac{X - \EX[X]}{\sqrt{\text{Var}[X]}} \leq x\right) - \Phi(x) \right\vert \leq \frac{0.3051\EX[|G_{\text{mcp}}|^3]}{(\EX[G_{\text{mcp}}^2])^{1.5}\sqrt{z}}
\end{equation*}
Since $G_{\text{mcp}}$ is an empirical distribution, the remaining terms can be calculated empirically to yield a final bound of
\begin{equation*}
    \sup_x \left\vert P\left(\frac{X - \EX[X]}{\sqrt{\text{Var}[X]}} \leq x\right) - \Phi(x) \right\vert \leq \frac{0.42228205813144853}{\sqrt{z}}
\end{equation*}

Similar to Appendix \ref{sec:poisson_to_gaussian}, a good upper-bound on the maximum CDF error found on the left-hand side is $0.01$, implying a theoretical error of at most 1\%. Solving, we find that $N$ must have a mean $z \geq 1784$. The real-world instrumentation that applications of this approximation aim to model often has systematic uncertainties (such as thermal noise, transient artifacts, shot noise, etc.) that frequently exceed 1\%. Therefore, chasing a tighter bound would demand drastically higher counts without improving the fidelity of the model.

Note that this formation of the bound provides a conservative, distribution-free guarantee because it controls the supremum norm of the CDF difference for all $x$. Further, the bound is driven by the third-order absolute moment which was empirically estimated. Empirical estimates of third-order absolute moments are known to heavily amplify the impact of tail observations. As such, the bound derived in this section should be treated as incredibly conservative.

\subsection{Mean of Final DN Image}
\label{sec:mean_final_dn}

This section derives $\EX[S(p, q)]$, whose formula is given in Equation \ref{eq:final_DN_equation}.
\begin{equation*}
    \EX\left[S(p, q)\right] = \EX\left[\sum_{k = 1}^{n_{\text{frame}}} S(p, q, k)\right]
\end{equation*}
\begin{equation*}
    = \EX\left[\sum_{k = 1}^{n_{\text{frame}}} \sum_{i=pn_{\text{bin}}}^ {(p+1)n_{\text{bin}} - 1} \sum_{j=qn_{\text{bin}}}^{(q+1)n_{\text{bin}} - 1} S(i, j, k)\right]
\end{equation*}
\begin{equation*}
    = \sum_{k = 1}^{n_{\text{frame}}} \sum_{i=pn_{\text{bin}}}^ {(p+1)n_{\text{bin}} - 1} \sum_{j=qn_{\text{bin}}}^{(q+1)n_{\text{bin}} - 1} \EX\left[S(i, j, k)\right]
\end{equation*}
\begin{equation*}
    = \sum_{k = 1}^{n_{\text{frame}}} \sum_{i=pn_{\text{bin}}}^ {(p+1)n_{\text{bin}} - 1} \sum_{j=qn_{\text{bin}}}^{(q+1)n_{\text{bin}} - 1} \EX\left[R(i, j, k)\right] + \EX\left[b(i, j, k)\right] + \EX\left[g_{\text{adc}}C(i, j, k)\right]
\end{equation*}
The term $C(i,j,k)$ has a different form depending on the location of $i,j$ (in the FOV vs. outside the FOV and etc.). This section focuses on the FOV case only, since the remaining cases are the same but with some terms dropped.
\begin{equation*}
    = \sum_{k = 1}^{n_{\text{frame}}} \sum_{i=pn_{\text{bin}}}^ {(p+1)n_{\text{bin}} - 1} \sum_{j=qn_{\text{bin}}}^{(q+1)n_{\text{bin}} - 1} b(i, j, k) + g_{\text{adc}}\EX\left[C_{\text{rad}}(k) + C_{\text{dark}}(k) + C_{\text{anom}}(i, j, k) + C_{\text{events}}(i, j, k)\right]
\end{equation*}
\begin{equation*}
    = \sum_{k = 1}^{n_{\text{frame}}} \sum_{i=pn_{\text{bin}}}^ {(p+1)n_{\text{bin}} - 1} \sum_{j=qn_{\text{bin}}}^{(q+1)n_{\text{bin}} - 1} b(i, j, k) + g_{\text{adc}}\left(c_{\text{rad}}(k) + c_{\text{dark}}(k) + c_{\text{anom}}(i, j, k) + \EX\left[C_{\text{events}}(i, j, k)\right]\right)
\end{equation*}
Use the results from Section \ref{sec:mean_counts}. Thus,
\begin{multline}
    \label{eq:mean_final_dn}
    \EX\left[S(p, q)\right] = \sum_{k = 1}^{n_{\text{frame}}} \sum_{i=pn_{\text{bin}}}^ {(p+1)n_{\text{bin}} - 1} \sum_{j=qn_{\text{bin}}}^{(q+1)n_{\text{bin}} - 1} b(i, j, k) + g_{\text{adc}}\\ \left(c_{\text{rad}}(k) + c_{\text{dark}}(k) + c_{\text{anom}}(i, j, k) + f_{\text{mcp}}(i, j) \EX[G_{\text{mcp}}] (e_{\text{photon}}(i,j,k) + e_{\text{rad}}(k))\right)
\end{multline}

\subsection{Variance of Final DN Image}
\label{sec:var_final_dn}

This section derives $\text{Var}[S(p, q)]$, whose formula is given in Equation \ref{eq:final_DN_equation}.
\begin{equation*}
    \text{Var}\left[S(p, q)\right] = \text{Var}\left[\sum_{k = 1}^{n_{\text{frame}}} S(p, q, k)\right]
\end{equation*}
\begin{equation*}
    = \text{Var}\left[\sum_{k = 1}^{n_{\text{frame}}} \sum_{i=pn_{\text{bin}}}^ {(p+1)n_{\text{bin}} - 1} \sum_{j=qn_{\text{bin}}}^{(q+1)n_{\text{bin}} - 1} S(i, j, k)\right]
\end{equation*}
Since each $S(i,j,k)$ is assumed to be independent from each other, the variance can be moved into summation.
\begin{equation*}
    = \sum_{k = 1}^{n_{\text{frame}}} \sum_{i=pn_{\text{bin}}}^ {(p+1)n_{\text{bin}} - 1} \sum_{j=qn_{\text{bin}}}^{(q+1)n_{\text{bin}} - 1} \text{Var}\left[S(i, j, k)\right]
\end{equation*}
\begin{equation*}
    = \sum_{k = 1}^{n_{\text{frame}}} \sum_{i=pn_{\text{bin}}}^ {(p+1)n_{\text{bin}} - 1} \sum_{j=qn_{\text{bin}}}^{(q+1)n_{\text{bin}} - 1} \text{Var}\left[R(i, j, k)\right] + \text{Var}\left[b(i, j, k)\right] + \text{Var}\left[g_{\text{adc}}C(i, j, k)\right]
\end{equation*}
The term $C(i,j,k)$ has a different form depending on the location of $i,j$ (in the FOV vs. outside the FOV and etc.). This section focuses on the FOV case only, since the remaining cases are the same but with some terms dropped.
\begin{equation*}
    = \sum_{k = 1}^{n_{\text{frame}}} \sum_{i=pn_{\text{bin}}}^ {(p+1)n_{\text{bin}} - 1} \sum_{j=qn_{\text{bin}}}^{(q+1)n_{\text{bin}} - 1} \text{Var}[R(i,j,k)] + g_{\text{adc}}^2\text{Var}\left[C_{\text{rad}}(k) + C_{\text{dark}}(k) + C_{\text{anom}}(i, j, k) + C_{\text{events}}(i, j, k)\right]
\end{equation*}
Each source of counts is independent and the first three sources of counts are assumed to be Poisson-distributed.
\begin{equation*}
    = \sum_{k = 1}^{n_{\text{frame}}} \sum_{i=pn_{\text{bin}}}^ {(p+1)n_{\text{bin}} - 1} \sum_{j=qn_{\text{bin}}}^{(q+1)n_{\text{bin}} - 1} \text{Var}[R(i,j,k)] + g_{\text{adc}}^2\left(c_{\text{rad}}(k) + c_{\text{dark}}(k) + c_{\text{anom}}(i, j, k) + \text{Var}\left[C_{\text{events}}(i, j, k)\right]\right)
\end{equation*}
Use the results from Section \ref{sec:var_counts}.
\begin{multline*}
    = \sum_{k = 1}^{n_{\text{frame}}} \sum_{i=pn_{\text{bin}}}^ {(p+1)n_{\text{bin}} - 1} \sum_{j=qn_{\text{bin}}}^{(q+1)n_{\text{bin}} - 1} \text{Var}[R(i,j,k)] + g_{\text{adc}}^2 \\ \left(c_{\text{rad}}(k) + c_{\text{dark}}(k) + c_{\text{anom}}(i, j, k) + f_{\text{mcp}}(i, j)^2\EX\left[G_{\text{mcp}}^2\right]\left(e_{\text{photon}}(i,j,k) + e_{\text{rad}}(k)\right)\right)
\end{multline*}

\end{document}